\documentclass[12pt]{article}
\usepackage{natbib}
\usepackage{graphicx}
\usepackage{array}
\usepackage{amsmath}
\usepackage{multirow}
\usepackage{amssymb}
\usepackage{xcolor}
\usepackage{url}
\usepackage[toc,page]{appendix}

\newcommand{\Esp}{\mathbb E}

\begin{document}

\title{Forecast score distributions with imperfect observations}

\title{Forecast score distributions with imperfect observations}
\author{Julie Bessac,$^{1}$ Philippe Naveau$^{2}$ \\
$^{1}$ {\small Mathematics and Computer Science Division, Argonne National Laboratory, Lemont, IL, 60439, USA} \\
$^{2}$ {\small Laboratoire de Sciences du Climat et de l'Environnement, IPSL-CNRS, Gif-sur-Yvette, 91191, France}}

\date{}

\maketitle

\begin{abstract}
The field of statistics has become one of the mathematical foundations in  forecast evaluations  studies, especially in regard to computing scoring rules.   
The  classical paradigm  of  scoring rules  is to  discriminate between two different forecasts by comparing them with observations.  
The probability distribution of the observed record is assumed to be perfect as a verification benchmark.   
In practice, however, observations are almost always  tainted by errors and uncertainties.    
These  may be due to homogenization problems, instrumental deficiencies, the need for indirect reconstructions from other sources (e.g., radar data),  model errors in gridded products like reanalysis,  or any other data-recording issues. 
If the yardstick used to compare forecasts is imprecise, one can wonder whether such types of errors may or may not have a strong  influence on decisions based on classical  scoring rules. 
We propose a new scoring rule scheme in the context of models that incorporate errors of the verification data. 
We rely on existing scoring rules and incorporate uncertainty and error of the verification data through a hidden variable and the conditional expectation of scores when they are viewed as a random variable. 
The proposed scoring framework is applied to standard setups, mainly an additive Gaussian  noise model and a multiplicative Gamma  noise model. 
These classical examples  provide  known and tractable conditional distributions and, consequently, allow us to interpret explicit expressions of our score. 
By considering scores as random variables one can access the entire range of their distribution.  
In particular we illustrate that the commonly used mean score can be a misleading representative of the distribution when this latter is highly skewed or have heavy tails. 
In a simulation study, through the power of a statistical test, we demonstrate the ability of the newly proposed score to better discriminate between forecasts when verification data are subject to uncertainty compared with the scores used in practice. 
We apply the benefit of accounting for the uncertainty of the verification data into the scoring procedure on a dataset of surface wind speed from measurements and numerical model outputs. 
Finally, we open some discussions to the use of this proposed scoring framework to  non-explicit  conditional distributions. 
\end{abstract}

\section{Introduction}\label{sec:intro}

\begin{figure}
\centering
\includegraphics[scale=.4]{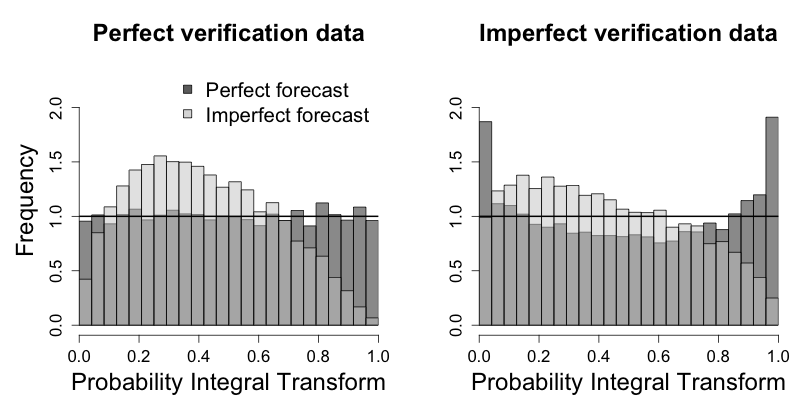}
\caption{An ``perfect'' (dark grey) and  ``imperfect'' (light grey) synthetic forecasts, with respective distributions $\mathcal{N}(0,2)$ and $\mathcal{N}(0.5,4)$, are compared with ``perfect'' verification data (left) with distribution $\mathcal{N}(0,2)$ and imperfect verification data $Y$ (right) with distribution $\mathcal{N}(0,3)$.  
Comparing an perfect forecast  (forecast having the same distribution as the true data) to corrupted verification data leads to an apparent under-dispersion of the perfect forecast.}
\label{fig:PIT}
\end{figure}

Probabilistic forecast evaluation generally involves the comparison of a probabilistic forecast cumulative distribution function (cdf) $F$ (in the following, we assume that $F$ admits a probability density function $f$) with verification data $y$ that could be of diverse origins \citep{Jolliffe04,Gneiting07b}.  
In this context verification data are most of the time, but not exclusively, observational data. 
Questions related to the quality and variability of observational data have been raised and tackled in different scientific contexts. 
For instance, data assimilation requires careful estimation of uncertainties related to both numerical models and observations \citep{Daley93,Waller14,Janjic17}.  
Apart from a few studies \citep[see, e.g.][]{hamill2006,Ferro17},  error and uncertainty associated with the verification data have rarely been addressed 
in forecast evaluation. 
Nonetheless, errors in verification data can lead to severe mis-characterization of probabilistic forecasts as illustrated in Figure \ref{fig:PIT}, where an perfect forecast (forecast having the same distribution as the true data) appears under-dispersed when the verification data have a larger variance than the true process. 
In the following, the underlying hidden process that is materialized by model simulations or measured through instruments, will be referred to as the true process. 
Forecast evaluation can be performed qualitatively through visual inspection of statistics of the data such as in Figure \ref{fig:PIT} 
 \cite[see, e.g.][]{Broecker20}, 
 and quantitatively through the use of scalar functions called scores \citep{Gneiting07b,Gneiting07}. 
In this work, we focus on the latter, we illustrate and propose a framework based on hidden variables to embed imperfect verification data into scoring functions via priors on the verification data and on the hidden state.  

\subsubsection*{Motivating example}
\begin{figure}
\centering
\includegraphics[scale=.4]{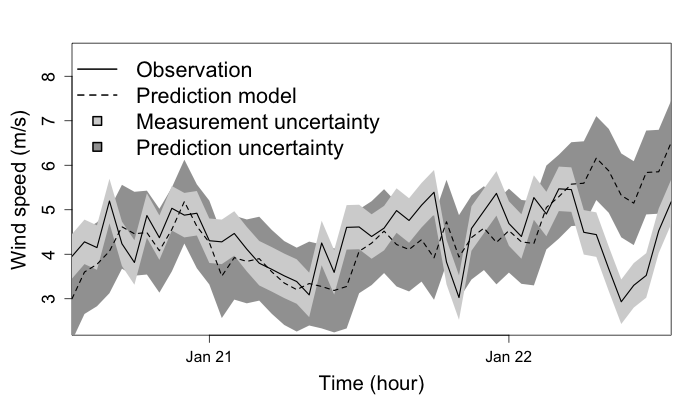}
\caption{Time series of surface wind speed measurements of two days of January 2012 in Wisconsin, USA. 
The solid line represents ground measurements and the light gray shaded area represents observational uncertainty ($\pm \sigma$ with $\sigma=0.5$ defined in  \cite{Pinson12}). 
The dashed line represents a numerical model output and the model uncertainty in dark gray shade. The model uncertainty refers to the standard deviation computed on the available forecast time series. }
\label{fig:timeseries_ws}
\end{figure}

To illustrate the proposed work, we consider surface wind data from a previous work \citep{bessac2018}.
Time series of ground measurements from the NOAA Automated Surface Observing System (ASOS) network are available at \url{ftp://ftp.ncdc.noaa.gov/pub/data/asos-onemin} and are extracted at 1 minute resolution. 
We focus on January 2012 in this study, the data are filtered via moving-average procedure and considered at an hourly level leading to $744$ data-points.
These ground-station data are considered as verification data in the following.  
Outputs from numerical weather prediction (NWP) forecasts are generated by using WRF v3.6 \citep{Skamarock08},  a state-of-the-art numerical weather prediction system designed to serve both operational forecasting and atmospheric research needs. 
The NWP forecasts are initialized by using the North American Regional Reanalysis fields dataset that covers the North American continent with a resolution of 10 minutes of a degree, 29 pressure levels (1000-100 hPa, excluding the surface), every 3 hours from the year 1979 until the present. Simulations are started every day during January 2012 with a forecast lead-time of 24 hours and cover the continental United States on a grid of 25x25 km with a time resolution of 10 minutes. 
Only one trajectory is simulated in this study. 
As observed in Figure \ref{fig:timeseries_ws}, the uncertainty associated with observation data can affect the evaluation of the forecast. 
As an extension, if  two forecasts were to fall within the uncertainty range of the observations, it would require a non-trivial  choice from the forecaster to rank forecasts. 
We will apply our proposed scoring framework to this dataset in Section \ref{sec:appli_wind} and compute scores embedding this uncertainty.

\subsubsection*{Imperfect verification data and underlying physical state}
In verification and recalibration of ensemble forecasts, an essential  verification  step is to find data that precisely identify the forecast events of interest, the so-called verification dataset \citep[see, e.g.,][]{Jolliffe04}.  
\cite{Jolliffe04}, in Section 1.3 of their book, discussed the uncertainty associated with verification data such as sampling uncertainty, direct measurement uncertainty, or changes in locations in the verification dataset. 
Verification data arise from various sources and consequently present various types of errors and uncertainties such as measurement error \citep{dirkson2019}, design and quality: e.g., gridded data versus weather stations, homogeneization problems, sub-sampling variability \citep{mittermaier2015}, or mis-matching resolutions. 
Resolutions mis-matching is becoming an issue due to weather or regional climate models running at ever finer resolution with increased fidelity for which observational data are rarely available for evaluation, hence requiring to account for up- or down-scaling error.   
In order to provide accurate forecast evaluations, it is crucial to distinguish and account for these types of errors.  
For instance, the effect of observational error can be stronger at short-term horizon when forecast error is smaller.   
Additive models have been used for the observational error \citep{ciach1999,saetra2004} or used to enhance coarser resolutions to allow comparison with fine resolution data \citep{mittermaier2015}. 
In the following, we will present an additive and a multiplicative framework based on a hidden variable to embed uncertainty in data sources. 

A common way of modeling errors is by representing the truth as a hidden underlying process also called state \citep{kalman1960,kalman1961}. 
Subsequently each source of data is seen as a proxy of the hidden state and modeled as a function of it. 
This forms the basis of data assimilation models where the desired state of the atmosphere is estimated through the knowledge of physics-based models and observational data that are both seen as versions of the non-observed true physical state.  
In the following, we base our scoring framework on the decomposition of the verification data as a function of a hidden true state,  referred to $X$, and an error term.

\subsubsection*{Related literature}
Uncertainty and errors arise from various sources such as the forecast and-or the verification data (predictability quality, uncertainty, errors, dependence and correlation, time-varying distribution) and the computational approximation of scores. 
A distribution-based verification approach was  initially proposed by \cite{Murphy87}, where joint distributions for forecast and observation account for the information and interaction of both datasets. 
\cite{wilks2010} studied the effect of serial correlation of forecasts and observations on the sampling distributions of Brier score and in particular serially correlated forecasts inflate its variance.  
\cite{bolin2019} discussed the misleading use of average scores, in particular for the continuous ranked probability score (CRPS) that is shown to be scale-dependent, when forecasts have varying predictability such as in non-stationary cases or exhibit outliers.  
\cite{brocker2007}, in their conclusion, highlighted the need of generalizing scores when verification data are uncertain in the context of properness and locality. 
More specifically, robustness and scale-invariance corrections of the CRPS are proposed to account for outliers and varying predictability in scoring schemes. 
Concerning the impact of the forecast density approximation from    finite ensembles, \cite{Zamo18}  
compared  four CRPS estimators   and highlighted recommendations in terms of the type of ensemble, whether random or a set of quantiles.  
In addition, several works focus on embedding verification data errors and uncertainties into scoring setups. 
Some methods aimed at  correcting  the verification data and use regular scoring metrics, such as perturbed ensemble methods \citep{Anderson96,Hamill01,Bowler08,Gorgas12}. 
Other works modeled observations as random variables and expressed scoring metrics in that context \citep{Candille08,Pappenberger09,Pinson12}. 
Some approaches directly modified the expression of metrics, \citep{hamill2006,Ferro17}, and others \cite[see, e.g.][]{hamill2006} accounted for varying climatology by sub-dividing the sample into stationary ones. 
In  \cite{ciach1999} and \cite{saetra2004}, additive errors were embedded in scores via convolution. 
 Analogously  to the Brier score decomposition of \cite{murphy1973}, \cite{Weijs11,weijs2010} decomposed   
the Kullback-Leibler divergence score and the cross-entropy score into uncertainty into  reliability and resolution components in order to account for uncertain verification data.
\cite{kleen2019} discussed scores scale sensitivity to additive measurement errors and proposed a measure of discrepancy between scores computed on uncorrupted and corrupted verification data.

\subsubsection*{Proposed scoring framework}
The following paper proposes an idealized framework to express commonly used scores with observational uncertainty and errors.  
The new framework relies on the decomposition of the verification data $y$ into a ``true'' hidden state  $x$ and an error term,  and on the representation of scores as a random variable when the verification data  is seen as a random variable.   
Information about the hidden state and the verification data are embedded via priors into the scoring framework, sharing analogies with  classical Bayesian analysis \citep{gelman2013}. 
More precisely, the proposed framework relies on the knowledge of the conditional distribution of the  ``true'' hidden state given the verification data, or on information which allows to calculate that conditional distribution. 

Section \ref{sec:score_correction} introduces a scoring framework that accounts for errors in the verification data for scores used in practice. 
Sections \ref{sec:additive_case}  and \ref{sec:multiplicative_case} respectively implement the additive and multiplicative error-embedding cases for the logarithmic score (log-score)  and CRPS. 
Section \ref{sec:illustrations} illustrates the merits of scores in a simulation context and a in real application case. 
Finally, Section \ref{sec:discussion} provides a final discussions and insights in future works. 
In particular,  we discuss the possible generalization of the proposed scoring framework when the decomposition of the verification data $y$ into a hidden state $x$ and an error does not fall into an additive or multiplicative setting as in Sections \ref{sec:additive_case}  and \ref{sec:multiplicative_case}.

\section{Scoring rules under uncertainty}\label{sec:score_correction}
In the following, we propose a version of scores based on the conditional expectation of what is defined an ``ideal'' score given the verification data tainted by errors, when scores are viewed as random variables. The idea of conditional expectation was also used in \citep{Ferro17}, but with a different conditioning and is discussed below as a comparison.

\subsection{Scores as random variables}
In order to build the proposed score version as well as to interpret scores further than through their mean and assess for their uncertainty, we will rely on scores represented as random variables. 
In practice, scores are averaged over all available verification data $y$, and the uncertainty associated with this averaged score is mostly neglected. 
However this uncertainty reveals to be significant as pointed it out in \cite{dirkson2019} where confident intervals were computed through bootstrapping.  
In order to assess, the probability distribution of score $s_{0}$, we assume $Y$ is a random variable representing the verification data $y$ and write a score as a random variable $s_{0}(F,Y)$, where $F$ is the forecast cdf to be evaluated.  
This representation gives access to the score distribution and enables to explore the uncertainty of this latter.  
A few works in the literature have already considered scores as random variables and performed associated analysis. 
In \cite{diebold2002} and \cite{jolliffe2007}, scores distributions were considered to build confidence intervals and hypothesis testing to assess differences in scores values when comparing forecasts to the climatology. 
In \cite{wilks2010}, the effect of serial correlation on the sampling distributions of Brier score was investigated. 
\cite{Pinson12} illustrated and discussed scores distributions across different prediction horizons and across different level of observational noise.

\subsection{Hidden state and scoring}
Scores used in practice are evaluated on verification data $y$, $s_{0}(F,y)$. 
We define the ``ideal'' score as $s_{0}(F,x)$, where $x$ is the realization of the hidden underlying ``true'' state that gives rise to the verification data $y$.   
Ideally, one would use $x$ to provide the best assessment of forecast $F$ quality through $s_{0}(F,x)$; however since $x$ is not available, we consider the best approximation of $s_{0}(F,x)$ given the observation $y$ in terms of $L^2$-norm via the conditional expectation. 
For a given score $s_0$, we propose the following score version $s_{\vee}(., y)$: 
\begin{equation}\label{eq:new_score}
s_{\vee}(F, y) =  \mathbb{E}\left(  s_0(F, X) |Y=y \right) 
\end{equation}
where $X$ is the true hidden state, $Y$ is the random variable representing the available observation used as verification data, and $F$ is the forecast cdf to be evaluated. 
One can view this scoring framework incorporating information about the discrepancy between the true state and the verification data in terms of errors and uncertainty in a Bayesian setting. 
To compute \eqref{eq:new_score}, we assume that the distributional features of the law of $X$, and the conditional law of $[Y | X]$, where $[.]$ and $[.|.]$ denotes respectively marginal and conditional distributions,  are directly available or can be obtained. 
This is  the classical framework used in data assimilation when both   observational and state equations are given, and the issue is infer the realization $x$   given observations $y$, see also our example in
Section \ref{sec:appli_wind}. 
Under this setup, the following properties hold for the score $s_{\vee}$: 
\begin{eqnarray}\label{eq:score_variance}
\nonumber \mathbb{E}_{X}\left[ s_0(F,X) \right]&=&\mathbb{E}_Y\left[ s_{\vee}(F, Y) \right]   \\
\mathbb{V}_{X}\left[ s_{0}(F, X) \right] &\geq& \mathbb{V}_{Y}\left[ s_{\vee}(F, Y) \right] \textrm{ for any forecast cdf $F$.} 
\end{eqnarray} 
Details of the computations are found in Appendix \ref{app:total_var}. 
The first equality guaranties that  any  propriety attached to the mean value of $s_{0}$ are preserved with  $s_{\vee}$.  
The second inequality arises from the law of total variance and implies a reduced dispersion of the corrected score compared to the ideal score.
This can be explained by the prior knowledge on the verification data that reduces  uncertainty in the score.   
These properties are illustrated with simulated examples and real data in Section \ref{sec:illustrations}.  

As a comparison, \cite{Ferro17} proposed a mathematical scheme to correct a score when error is present in the verification data $y$.  
Ferro's modified score, denoted  $s_{F}(f, y)$ in this work,  is derived from a classical score, say  $s_0(f, x)$ where $x$ is the hidden true state.   
With these notations,  the corrected score $s_{F}(f, y)$ is built such that it respects  the following conditional expectation
$$s_0(f, x) =  \mathbb{E}\left(  s_{F}(f, Y) |X=x \right).$$  
In other words, the score $ s_{F}(f, y)$ computed from the $y$'s provides the same value on average that the proper score computed from the unobserved true state $x$'s. 
The modified score $s_F$ explicitly targets biases in mean induced by imperfect verification data. 
In terms of assumptions, we note that  the conditional law of $Y$ given $X$ needs to be known in order to compute 
$s_0(f, x)$ from  $s_{F}(f, y)$, e.g. see Definitions 2 and 3 in \cite{Ferro17}. 
In particular, in \cite{Ferro17} the correction framework  is illustrated with the logarithmic score in the case of a Gaussian error model (i.e. $[Y|X=x] \sim \mathcal{N}(x,\omega^2)$).

\subsubsection*{Implementation and generalization} 
The proposed score reveals desirable mathematical properties of unbiasedness and variance reduction while accounting for the error in the verification data; however it relies on the knowledge of  $[X|Y]$,  or equivalently of $[Y|X]$ and $[X]$. 
We assume that the decomposition of $Y$ into  a hidden state $X$ and an error, is given and is fixed with respect of the evaluation framework.
Additionally, the new framework relies on the necessity to integrate $s_{0}(f,x)$ against conditional distributions, which might require some quadrature approximations in practice when closed formulations are not available. 
In this work, we assume to have access to the climatology distribution and we rely on this assumption to compute the distribution of $X$ or priors of is distribution. However, depending on the context and as exemplified in Section \ref{sec:appli_wind}, alternative definitions and computations of $X$ can be considered as for instance relying on known measurement error models.  
Nonetheless, as illustrated in Sections \ref{sec:additive_case} and \ref{sec:multiplicative_case}, the simplifying key to the score derivation in Eqn. \eqref{eq:new_score} is to use Bayesian conjugacy when applicable.  
This is illustrated in the following sections with a Gaussian additive case and a Gamma multiplicative case. 
Although the Bayesian conjugacy is a simplifying assumption, as in most Bayesian settings it is not a necessary condition and for cases with non explicit conjugate priors, all Bayesian and/or  assimilation tools (e.g., via sampling algorithms such as Markov Chain Monte Carlo methods \citep{robert2013} or non-parametric approaches such as Dirichlet processes) could be used to draw samples from $[ X| Y]$ and estimate  the distribution  $[s_0(F, X) |Y=y]$  for a given $s_0(F,.)$.
Finally, in the following we assume that distributions have known parameters; however this assumption can be relaxed by considering prior distributions on each involved parameter, via  hierarchical Bayesian modeling. 
The scope of this paper is to provide a conceptual tool and the question of parameters estimation, besides the example in Section \ref{sec:illustrations}, is not treated in detail. 
In Section \ref{sec:illustrations}, we apply the proposed score derivation to the aforementioned  surface wind data example described in the introduction.     
In Section  \ref{sec:discussion}, we discuss the challenges of computing scores as defined in  Eqn. \eqref{eq:new_score} or in \cite{Ferro17} in more general cases of state-space models that are not limited to additive or multiplicative cases.

\section{Gaussian additive case}\label{sec:additive_case} 
As discussed earlier, a commonly used setup in applications is when errors are additive and their natural companion distribution is Gaussian. 
Hereafter, we derive the score from Eqn. \eqref{eq:new_score} for the commonly used log-score  and CRPS in the Gaussian additive case, where the hidden state $X$ and the verification data $Y$ are linked through the following system: 
\begin{equation*}
\mbox{ Model (A)} 
\; \left\{ 
	\begin{array}{ll}
		X \sim \mathcal{N}(\mu_0,\sigma^2_0), \\
		Y = X   +   \mathcal{N}(0,\omega^2),\\
	\end{array}
	\right.
\end{equation*}
where $Y$ is the observed verification data, $X$ is the hidden true state, and all Gaussian variables are assumed independent. 
In the following, for simplicity we consider $\Esp(X)=\Esp(Y)$; however one can update the model easily to a mean-biased verification data $Y$. 
Parameters are supposed to be known from the applications, one could use priors on the parameters when estimates are not available. 
In the following, we express different versions of the log-score and CRPS: the ideal version, the used-in-practise version, and the error-embedding version from Eqn. \eqref{eq:new_score}. 
In this case, as well as in Section \ref{sec:multiplicative_case}, since conditional distributions are expressed through Bayesian conjugacy, most computational efforts rely on integrating the scores against the conditional distributions. 

\subsection{Log-score versions}
For a Gaussian predictive pdf $f$ with mean $\mu$ and variance $\sigma^2$, the  log-score is defined by 
\begin{equation}\label{eq:logscore_gauss}
s_0(f, x) = \log \sigma + \frac{1}{2 \sigma^2}(x-\mu)^2 + \frac{1}{2}\log 2 \pi,
\end{equation}
has been widely used in the literature.  
Ideally, one would access the true state $X$ and evaluate forecast against its realizations $x$; however since $X$ is not accessible scores are computed against observations $y$: 
\begin{equation}\label{eq:logscore_gaussY}
s_0(f, y) = \log \sigma + \frac{1}{2 \sigma^2}(y-\mu)^2 + \frac{1}{2}\log 2 \pi.
\end{equation}
Applying  basic properties of Gaussian conditioning, our score defined by \eqref{eq:new_score} can be written as: 
\begin{eqnarray}\label{eq:new_logscore_gauss}
s_{\vee}(f, y) & = & \log \sigma + \frac{1}{2 \sigma^2}   \left\{\frac{\omega^2\sigma_0^2}{\sigma_0^2+\omega^2}  +  \left( \bar{y} -\mu \right)^2 \right\}+ \frac{1}{2}\log 2 \pi \\
\nonumber && \textrm{where } \bar{y} =\frac{\omega^2}{\sigma_0^2+\omega^2} \mu_0 +  \frac{\sigma_0^2}{\sigma_0^2+\omega^2} y.
\end{eqnarray}
In particular, $\bar{y}=\Esp(X|Y=y)$ arises from the conditional expectation of $s_(f,X)$ in \eqref{eq:logscore_gauss} and is a weighted sum that updates the prior information about $X\sim \mathcal{N}(\mu_0,\sigma^2_0)$ with the observation $Y \sim  \mathcal{N}(\mu_0, \sigma^2_0 + \omega^2)$. 
In the following equations,  $\bar{y}$ represents the same  quantity.  
Details of the computations are found in Appendix \ref{app:cond_logscore} and rely on the integration of  Eqn. \eqref{eq:logscore_gauss} against the conditional distribution of $[X|Y=y]$. 

As a comparison, the corrected score from \cite{Ferro17} expresses as $\displaystyle s_{F}(f, y) = \log \sigma + \frac{(y-\mu)^2 - \omega^2}{2 \sigma^2} + \frac{1}{2}\log 2 \pi$ with the same notations and under the assumption $[Y|X=x]\sim\mathcal{N}(x,\omega^2)$. 
In this case, the distribution of the hidden state $x$ is not involved; however $y$ is not scale-corrected, only a location-correction is applied compared to Eqn. \eqref{eq:new_logscore_gauss}.

\subsection{CRPS versions}
Besides the logarithmic score, the CRPS is another classical  scoring rule used in weather forecast centers.
It is defined as
\begin{equation}\label{eq:crps}
c_0(f,x) = \mathbb{E} |Z-x| -\frac{1}{2} \mathbb{E} |Z-Z'|, 
\end{equation}
where $Z$ and $Z'$ are {\it i.i.d.} copy random variables with continuous  pdf $f$. 
The CRPS can be rewritten as 
$c_0(f,x) =x + 2   \mathbb{E}( Z-x)_+ - 2 \mathbb{E}(Z\overline{F}(Z)),$
where $( Z-x)_+$ represents the positive part of $Z-x$ 
and $\overline{F}(x)=1-F(x)$ corresponds to the survival function associated to the cdf $F$. 
For example, the CRPS for a Gaussian forecast with parameters $\mu$ and  $\sigma$  is equal to
\begin{equation}\label{eq:CRPS_Gauss}
c_0(f,x) = x + 2 \sigma \left[  \phi\left( \frac{x-\mu}{\sigma}\right) -   \frac{x-\mu}{\sigma}\overline{\Phi}\left( \frac{x-\mu}{\sigma}\right)     \right]  - \left[ \mu  + \frac{\sigma}{\sqrt{\pi}}     \right] ,
\end{equation}
where $\phi$ and $\Phi$ are the pdf and cdf of a standard normal distribution  \citep{Gneiting05,Taillardat16}. 
Similarly to \eqref{eq:logscore_gaussY}, in practice one evaluates \eqref{eq:CRPS_Gauss} against observations $y$ since the hidden state $X$ is unobserved.

Under the Gaussian additive model (A), the proposed CRPS defined by Eqn. \eqref{eq:new_score} is written as 
\begin{equation}\label{eq:CRPS_corr}
 c_{\vee}\left(f,y\right) =   \bar{y}   + 
 2 \sigma_{\omega}  \left[ \phi\left( \frac{\bar{y}-\mu}{\sigma_{\omega}}\right) - \frac{\bar{y}-\mu}{\sigma_{\omega}} \bar{\Phi} \left( \frac{\bar{y}-\mu}{\sigma_{\omega}}\right) \right] - \left[ \mu + \frac{\sigma}{\sqrt{\pi}}\right],
 \end{equation}
where 
$\sigma^2_{\omega}=\sigma^{2}+\frac{\omega^2 \sigma_0^2}{\sigma_0^2+\omega^2}$
and
$\bar{y}=\frac{\omega^2}{\sigma_0^2+\omega^2} \mu_0 +  \frac{\sigma_0^2}{\sigma_0^2+\omega^2}$ as defined above.   
Details of the computations are found in Appendix \ref{app:cond_crps} and rely on similar principles as the above paragraph.

\subsection{Scores distributions}\label{sec:score_distrib}
\begin{figure}
\centering
\includegraphics[scale=.44]{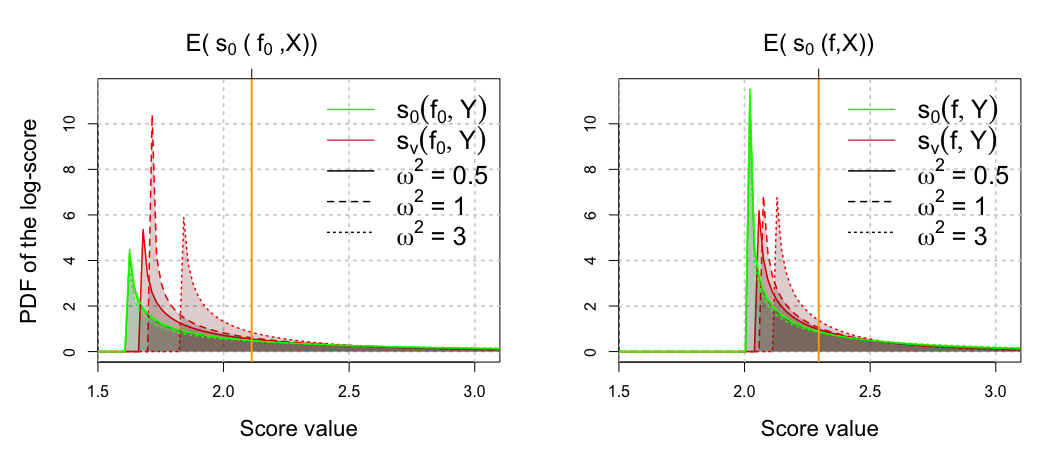}
\caption{Probability distribution functions (pdf) from Eqn. \eqref{eq:logSc_distrib} of the log-score used in practice $s_{0}(.,Y)$ (green) and the proposed score $s_{\vee}(.,Y)$ (red) computed for the perfect forecast $f_0$ on the left ($\mu=\mu_0=0$ and $\sigma=\sigma_{0}=2$)  and imperfect forecasts $f$ on the right ($\mu=1$ and $\sigma=3$). The mean of the ideal score is depicted with an orange line: $\Esp(s_{0}(f_{0},X))$ on the left and  $\Esp(s_{0}(f,X))$ on the right. The following distributions are used: $X\sim \mathcal{N}(0,4)$ and $Y\sim \mathcal{N}(0, 4 + \omega^{2})$ with several levels of observational noise $\omega^2 = 0.5, 1, 3$. } \label{fig:density_logSc}
\end{figure}

Under the Gaussian additive model (A), the random variables associated with the proposed log-scores defined by \eqref{eq:logscore_gaussY} and \eqref{eq:new_logscore_gauss}
is written as 
\begin{equation}\label{eq:logSc_distrib}
s_{0}\left(f, Y\right) \stackrel{d}{=}   a_{0}+   b_{0} \chi^2_{0}, 
 \mbox{\hspace{.25cm}  and  \hspace{.25cm}} 
s_{\vee}\left(f, Y\right) \stackrel{d}{=}  a_{\vee}  +    b_{\vee}  \chi^2_{\vee},
\end{equation}
where $\stackrel{d}{=} $ means equality in distribution and 
$ \chi^2_{0}$ and $ \chi^2_{\vee}$ noncentral chi-squared random variables with {\it 1 d.o.f.} and
respective non-centrality parameters $\lambda_{0}$ and $\lambda_{\vee}$. 
The explicit expressions of the constants $\lambda_{0}$, $\lambda_{\vee}$ $a_{0}$, $a_{\vee}$, $b_{0}$ and  $b_{\vee}$ are found in Appendix \ref{app:logsc_distrib}. 
The distribution of $s_{0}\left(f, X\right)$ can be derived similarly. 
Figure \ref{fig:density_logSc} illustrates the distributions in \eqref{eq:logSc_distrib} for various values of the noise parameter $\omega$. 
The distributions are very peaked due to the single degree of freedom of the Chi-square distribution, moreover their bulks are far from the true mean of the ideal score of $s_{0}(.,X)$ challenging the use of the mean score to compare forecasts. 
The concept of propriety is based on averaging scores;  however the asymmetry and long right tails of the noncentral chi-squared densities makes the mean a non-reliable statistic to represent such distributions. 
\cite{bolin2019} discussed the misleading use of averaged scores in the context of time-varying predictability where different scales of prediction errors arise generating different of order of magnitude of evaluated scores. 
However, the newly proposed scores exhibit a bulk of their distribution closer to the mean and with a reduced variance as stated in Eqn. \eqref{eq:score_variance}, leading to more confidence in the mean when this latter is consider. 

Similarly, under the  additive Gaussian model (A), the random variable associated with the proposed CRPS defined by (\ref{eq:CRPS_corr}) is written as 
\begin{equation}\label{eq:CRPS_corr_dist}
 c_{\vee}\left(f,Y\right) =   \bar{Y}   + 
 2 \sigma_{\omega}  \left[ \phi\left( \frac{\bar{Y}-\mu}{\sigma_{\omega}}\right) - \frac{\bar{Y}-\mu}{\sigma_{\omega}} \bar{\Phi} \left( \frac{\bar{Y}-\mu}{\sigma_{\omega}}\right) \right] - \left[ \mu + \frac{\sigma}{\sqrt{\pi}}\right],
 \end{equation}
where 
$\sigma^2_{\omega}=\sigma^{2}+\frac{\omega^2 \sigma_0^2}{\sigma_0^2+\omega^2}$
and the random variable $\bar{Y}=\frac{\omega^2}{\sigma_0^2+\omega^2} \mu_0 +  \frac{\sigma_0^2}{\sigma_0^2+\omega^2} Y$ follows a Gaussian pdf with mean $\mu_0$ and variance $\sigma_0^2 \times \frac{\sigma_0^2}{\sigma_0^2+\omega^2}$. 
The distribution of \eqref{eq:CRPS_corr_dist} does not belong to any known parametric families; however it is still possible to characterize the score distribution through sampling when the distribution of $Y$ is available.   
Finally, having access to distributions like \eqref{eq:logSc_distrib} or \eqref{eq:CRPS_corr_dist} gives access to the whole range of uncertainty of the score distributions helping to derive statistics that are more representative than the mean as pointed out above, and to  compute confidence intervals without bootstrapping approximations as in \citep{wilks2010,dirkson2019}.
Finding adequate representatives of a score distribution that shows reliable discriminative skills is beyond the scope of this work. 
Nevertheless, in Appendix \ref{sec:score_wasserstein} we take forward the concept of score distributions and apply it to computing distances between score distributions in order to assess their discriminative skills.

\section{Multiplicative Gamma case}\label{sec:multiplicative_case}
The Gaussian assumption is appropriate when dealing with averages, for example, mean temperatures; however, the
normal hypothesis cannot be justified for positive and skewed variables such as precipitation intensities.  
For instance, multiplicative models are often used in hydrology to account for various errors and uncertainty (measurement errors, process uncertainty and unknown quantities), see \citep{kavetski1,kavetski2,mcmillan2011rainfall}. 
An often-used alternative in such cases is to use a Gamma distribution, which works fairly well in practice to represent the bulk of rainfall intensities.  
Hence, we assume in this section that the true but unobserved $X$ follows a Gamma distribution with parameters $\alpha_0$ and $\beta_0$: 
$\displaystyle f_{X}(x)= \frac{\beta_0^{\alpha_0}}{\Gamma(\alpha_0)} x^{\alpha_0-1} \exp( - \beta_0 x)$, for $x >0$.  
For positive random variables such as precipitation,  additive models cannot be used to introduce errors. 
Instead, we prefer to use a multiplicative model of the type
\begin{equation}
\mbox{ Model (B)} 
\; \left\{ 
	\begin{array}{ll}
		X \sim \mbox{Gamma}(\alpha_0,\beta_0), \\
		Y = X   \times \epsilon,\\
	\end{array}
	\right.
\end{equation}
where $\epsilon$ is a positive random variable independent of $X$. 
To make feasible computations, we model the error $\epsilon$ as an inverse Gamma pdf with parameters $a$ and $b$: 
$\displaystyle f_{\epsilon}(u)= \frac{b^a}{\Gamma(a)} u^{-a-1} \exp\left( - \frac{b}{u}\right),$ for $u >0$.
The basic conjugate prior properties of such Gamma and inverse Gamma distributions  allows us to  easily derive the pdf $[X|Y=y]$. 
Analogously to Section \ref{sec:additive_case}, we express the  log-score and  CRPS within this multiplicative Gamma model in the following paragraphs.

\subsection{Log-score versions}
Let us consider a Gamma distribution $f$ with parameters $\alpha>0$ and $\beta>0$ for the prediction.  
With obvious notations, the log-score for this forecast $f$ becomes 
\begin{equation}\label{eq:logScGamma}
s_{0}(f,x) = (1-\alpha) \log x + \beta x - \alpha \log\beta + \log \Gamma(\alpha). 
\end{equation}
Under the Gamma multiplicative model (B), the random variable associated with  the corrected log-scores defined by Eqn. \eqref{eq:new_score} and  (\ref{eq:logScGamma})
is expressed as 
\begin{equation}\label{eq:logScCorrGamma}
s_{\vee}(f, Y) = (1-\alpha) \left( \psi(\alpha_0+a)-\log (\beta_0+b/Y) \right)  + \beta \frac{\alpha_0+a}{\beta_0+b/Y} - \alpha \log\beta + \log \Gamma(\alpha),
\end{equation}
where $\psi(x)$ represents the digamma function defined as the logarithmic derivative of the Gamma function, namely, $\psi(x)=d \log \Gamma(x)/dx$. 
Details of the computations are found in Appendix \ref{app:logsc_mult}.

\subsection{CRPS versions} 
\begin{figure}
\centering
\includegraphics[scale=.42]{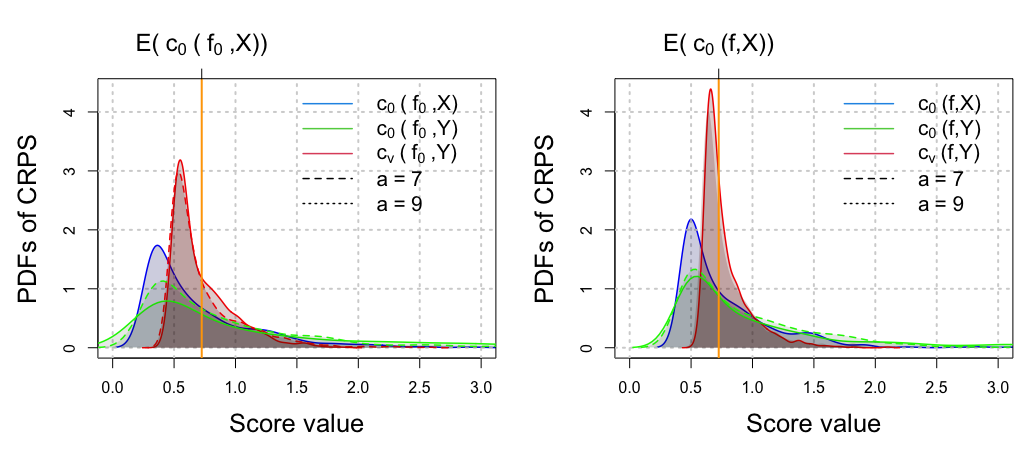}
\caption{Estimated probability distribution of the CRPS under the multiplicative Gamma model; shown in blue is the distribution of $c_{0}(.,X)$; shown in green is the  distribution of $s_{0}(.,Y)$; and shown in red is the distribution of $c_{\vee}(., Y)$ respectively from Equations \eqref{eq:crps_Gamma} and \eqref{eq:crps_Gamma_corr}. 
The mean of the ideal score $\Esp(c_{0}(.,X))$ is depicted with an orange line. 
Left: score distributions for the perfect forecast $f_0$ having the same distribution as $X$ ($\alpha=\alpha_{0}=7$ and $\beta=\beta_{0}=2$) validated against $X$ and corrupted verification data $Y$ with different levels of error: $a=7,9$ and $b=8$. 
Right:  score distributions for the imperfect forecast $f$ with distribution parameters $\alpha=4$ and $\beta=1$ validated against $X$ and corrupted verification data $Y$. 
The following parameters $\alpha_{0}=7$, $\beta_{0}=2$ are used for the hidden state $X$.}\label{fig:crps_gamma}
\end{figure}

For a Gamma forecast with parameters $\alpha$ and  $\beta$,  the corresponding CRPS  \citep[see, e.g.,][]{Taillardat16,Scheuerer15} is equal to  
\begin{equation}\label{eq:crps_Gamma}
c_0(f,x)  =  \left[ \frac{\alpha}{\beta}   -  \frac{1}{\beta B(.5,\alpha)}   \right] - x + 2 \left[ \frac{x}{\beta} f\left( x\right) +  \left(\frac{\alpha}{\beta}-x\right)\overline{F}\left( x\right)     \right].
\end{equation}
Under the multiplicative Gamma model (B), the random variable associated with the CRPS \eqref{eq:crps_Gamma} corrected by Eqn. \eqref{eq:new_score} is expressed as 
\begin{eqnarray}\label{eq:crps_Gamma_corr}
\nonumber c_{\vee}(f, Y) & = & \left[ \frac{\alpha}{\beta}   -  \frac{1}{\beta B(.5,\alpha)}   \right] -  \frac{\alpha_{0}+a}{\beta_{0}+\frac{b}{Y}} +  2\frac{\beta^{\alpha-1}(\beta_{0}+b/Y)^{\alpha_{0}+a}}{B(\alpha,\alpha_{0}+a)(\beta+ \beta_{0}  + b/Y)^{\alpha+\alpha_{0} + a}} \\
\nonumber  &  & + \frac{2(\beta_{0}+b/Y)^{\alpha_{0}+a}}{\Gamma(\alpha)\Gamma(\alpha_{0}+a)}\int_{0}^{+\infty} \left( \frac{\alpha}{\beta} - x\right)\Gamma(\alpha,\beta x) x^{\alpha_{0}+a-1} \exp(-(\beta_{0}+b/Y)x) \mathrm{d}x. \\
\end{eqnarray}
Details of the computations are found in Appendix \ref{app:crps_mult}.  
Similarly to Section \ref{sec:score_distrib}, one can access the range of uncertainty of the proposed scores \eqref{eq:logScCorrGamma} and \eqref{eq:crps_Gamma_corr} when sampling from the distribution of $Y$ is available.   
As an illustration, Figure \ref{fig:crps_gamma} shows the distributions of the three CRPS presented in this section. 
Similarly to the previous section, one can see the benefits of embedding the uncertainty of the verification data are noticeable in the variance reduction of distributions shaded in red and the smaller distance between the bulk of distributions in red and the mean value of the ideal score.  
%

\section{Applications and illustrations}\label{sec:illustrations}

The following section applies and illustrates through simulation studies the benefit of accounting for uncertainty in scores as presented in Section \ref{sec:score_correction} through the power analysis of a hypothesis test and the  numerical illustration of  the motivating example with wind data from Section \ref{sec:intro}.  
In Appendix \ref{sec:score_wasserstein}, we illustrate further the consideration of score distributions via the Wasserstein distance.

\subsection{Power analysis of hypothesis testing}
In this section, a power analysis of the following hypothesis test is performed on simulated data following \cite{diebold2002} in order to investigate the discriminative skills of the proposed score from Eqn. \eqref{eq:new_score}.   
In \cite{diebold2002}, hypothesis tests focused on discriminating forecasts from the climatology; in the current study, we test the ability of the proposed score to discriminate any forecast $f$ from the perfect forecast $f_{0}$ (forecast having the ``true'' hidden state $X$ distribution).  
We consider the reference score value of the perfect forecast $f_{0}$ being the  mean score evaluated against the true process $\Esp(s_{0}(f_{0},X))$.  
Hypothesis tests are tuned to primarily minimize the error of type I (wrongly rejecting the null hypothesis), consequently it is common practice to assess the power of a test.  
The power $p$ is the probability of rejecting a false null hypothesis, the power expresses as $1-\beta$ where $\beta$ is the error of type II (failing to reject a false null hypothesis) and expresses as 
$$p = P(\mbox{Rejecting } \mathrm{H}_{0} | \mathrm{H}_{1}\mbox{ true}).$$
The closer to $1$ the power is, the better the test is a detecting a false null hypothesis.  
For a given forecast $f$, the considered hypothesis are expressed as 
\begin{equation}\label{eq:test}
\left\{  \begin{array}{l}
\mathrm{H}_{0}: \mbox{Forecast $f$ is perfect ($f=f_{0}$) through the score $s$ leading to } \Esp(s(f,.)) \mbox{ close to } \Esp(s_{0}(f_{0},X)),  \\
\mathrm{H}_{1}: \mbox{Forecast $f$ is imperfect leading to } \Esp(s(f,.)) \mbox{ far from } \Esp(s_{0}(f_{0},X)) 
\end{array}\right.
\end{equation}
where $f_0$ and $f$ are respectively the perfect forecast  and an imperfect forecast to be compared with the perfect forecast $f_0$. 
In the following, the score $s(f,.)$ will represent respectively $s_{0}(f, X)$, $s_{0}(f, Y)$ and $s_{\vee}(f, Y)$ in order to compare the ability of the ideal score, the score used in practice and the proposed score. 
The parameters associated with the hypothesis are the parameters, $\mu$ and $\sigma$ in the following additive Gaussian model, of the imperfect forecast $f$ and they are varied to compute the different powers.  
The statistical test corresponding to \eqref{eq:test} expresses as 
\begin{equation}\label{eq:testbis}
\left\{  \begin{array}{l}
\mathrm{H}_{0} \mbox{ is accepted if } t \leq c  \\
\mathrm{H}_{0} \mbox{ is rejected if } t>c
\end{array}\right.
\end{equation}
where $t=|\Esp(s(f,.)) - \Esp(s_{0}(f_{0},X))|$ is the test statistics and $c$ is defined via the error of type I $P(t > c|H_{0}\mbox{ is true})=\alpha$ with the level $\alpha=0.05$ in the present study. 

To illustrate this test with numerical simulations, the additive Gaussian model (A) is considered for the log-score,  where the forecast, $X$ and $Y$ are assumed to be normally distributed with an additive error. 
The expectation in \eqref{eq:testbis} is approximated and computed over a $N=1000$ verification data-points, being the true state $X$ and the corrupted data $Y$. 
The approximated test statistic is denoted with $\hat{t}$. 
The power $p$ expresses as $P(\hat{t} > c | f \mbox{ is imperfect})$ and is computed over $10000$ samples of length $N$ when parameters $\mu$ and $\sigma$ of the forecast $f$ are varied.

\begin{figure}
\centering
\includegraphics[scale=.23]{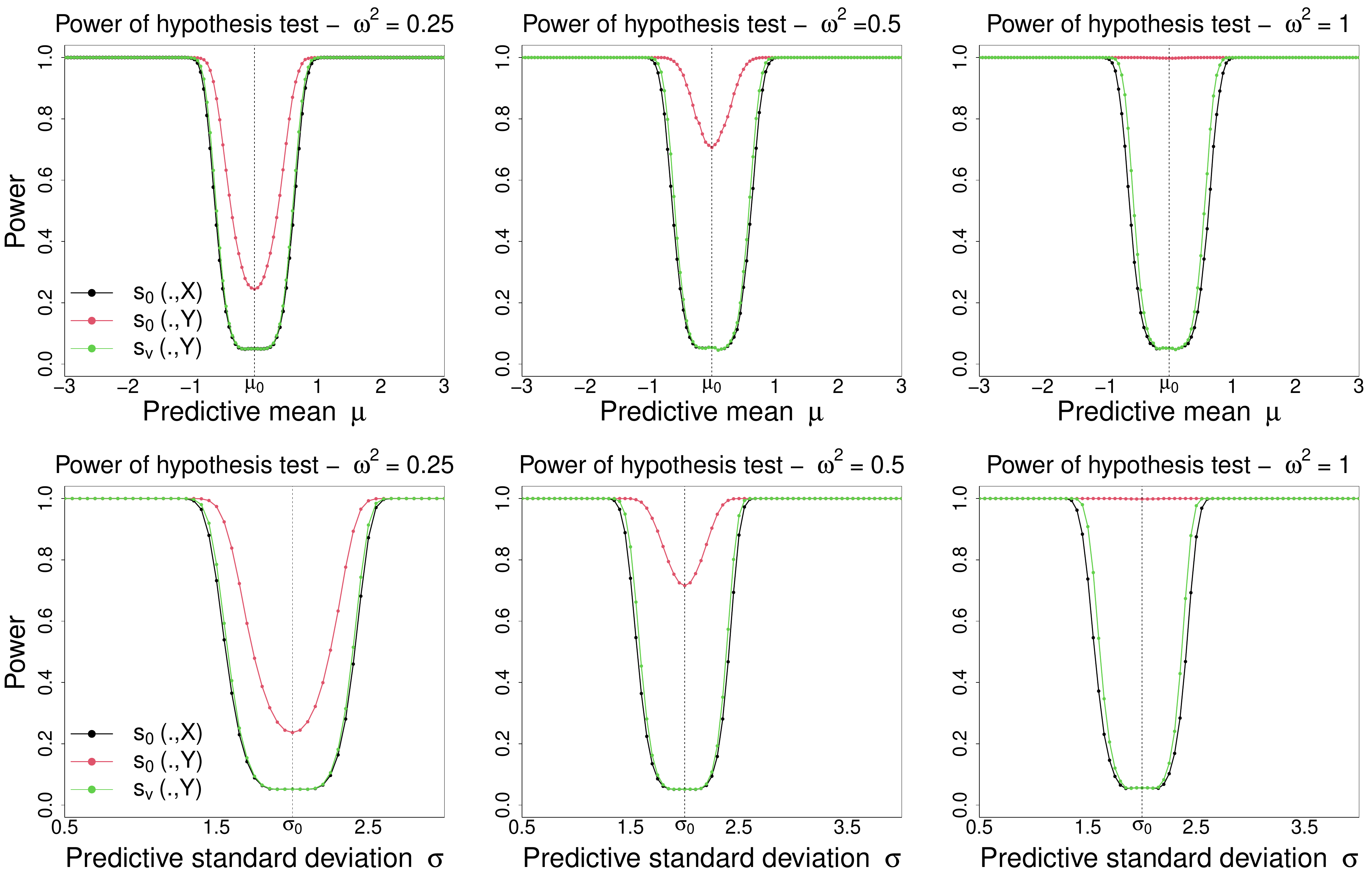}
\caption{Power of the test \eqref{eq:testbis} against varying predictive mean $\mu$ (top) and against varying predictive standard deviation $\sigma$ (bottom) of the forecast $f$ for different observational noise levels $\omega^2=0.25$ (left), $\omega^2=0.5$ (central) and $\omega^2=1$ (right). 
In the simulations, the true state $X$ is distributed as $\mathcal{N}(\mu_{0}=0,\sigma_{0}^{2}=4)$ and $Y$ is distributed as $\mathcal{N}(0,\sigma_{0}^{2}+\omega^{2})$.  
The power expected to be low around $\mu_0$ (resp. $\sigma_{0}$) and as large as possible elsewhere. }\label{fig:powerVSmu}
\end{figure}

In Figure \ref{fig:powerVSmu}, the above power $p$ is shown for varying mean $\mu$ and standard deviation $\sigma$ of the forecast $f$ in order to demonstrate the ability of the proposed score $s_{\vee}(.,Y)$ to better discriminate between forecasts than the commonly used score $s_{0}(.,Y)$.    
One expects the power to be low around respectively $\mu_{0}$ and $\sigma_{0}$, and as high as possible outside these values.  
We notice that the ideal score $s_{0}$ and the proposed score $s_{\vee}$ have similar power for the test \eqref{eq:testbis} suggesting similar discriminating skills for both scores.  
However, the commonly used in practice score $s_{0}(.,Y)$ results in an ill-behaved power as the observational noise increases (from left to right), indicating the necessity to account for the uncertainty associated with the verification data.  
The ill-behaved power illustrates the propensity of the test based on $s_{0}(.,Y)$ to reject $H_0$ too often and in particular wrongfully when the forecast $f$ is perfect and equals $f_0$. 
In addition, the power associated with the score $s_{0}(.,Y)$ fails to reach the nominal test level $\alpha$ due to the difference in means between $s_{0}(.,X)$ and $s_{0}(.,Y)$ ($\Esp(s_{0}(f,X)) \neq \Esp(s_{0}(f,Y))$ for any forecast $f$) caused by the errors in the verification data $Y$. This highlights the unreliability of scores evaluated against corrupted verification data.  
Both varying mean and standard deviation reveal similar conclusions regarding the ability of the proposed score to improve its discriminative skills over a score evaluated on corrupted evaluation data.  

\subsection{Application to wind speed prediction}\label{sec:appli_wind}
\begin{figure}
\centering
\includegraphics[scale=.42]{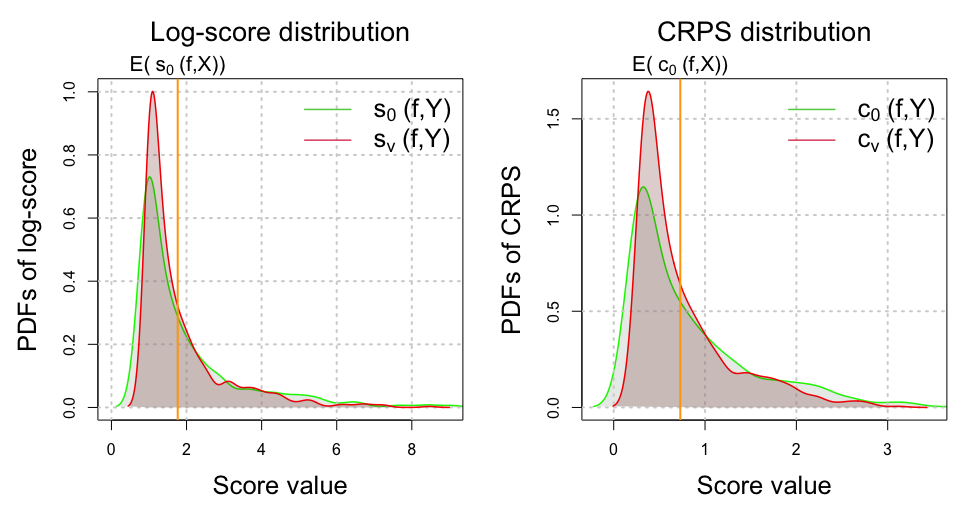}
\caption{Empirical distribution of uncorrected (green) and corrected scores (red) for the log-score (left) and CRPS (right) for the probabilistic distribution NWP model 1 evaluated against observations tainted with uncertainty.  
The ideal mean score value is illustrated by the orange vertical line. }\label{fig:distrib_score_wind}
\end{figure}

As discussed in the motivating example of Section \ref{sec:intro}, we consider surface wind speed data from  the previous work \citep{bessac2018} and associated probabilistic distributions.   
In \cite{bessac2018}, a joint distribution for observations, denoted here $X_{ref}$, and NWP model outputs is proposed and based on Gaussian distributions in a space-time fashion.  
This joint model aimed at predicting surface wind speed based on the conditional distribution of $X_{ref}$ given NWP model outputs.  
The data are Box-Cox-transformed in this study to approximate normal distributions.  
The model was fitted by maximum likelihood for an area covering parts of Wisconsin, Illinois, Indiana and Michigan in the United States.  
In this study, we focus on one station in Wisconsin and recover the parameters of its marginal joint distribution of observations $X_{ref}$ and NWP outputs from the joint spatio-temporal model.   
In the following, we evaluate with scores the probability distribution of the NWP model outputs depicted in Figure \ref{fig:timeseries_ws}.  
In \cite{bessac2018}, the target distribution was the fitted distribution of the observations $X_{ref}$;  
however in the validation step of the predictive probabilistic model, the observations shown in  Figure \ref{fig:timeseries_ws} were used without accounting of their potential uncertainty and error.  
This leads to a discrepancy between the target variable $X_{ref}$ and the verification data that we denote as $Y_{obs}$.  
From \cite{Pinson12}, a reasonable model for unbiased measurement error in wind speed is $\epsilon_{obs} \sim \mathcal{N}(0,0.25)$.  
Subsequently to Section \ref{sec:additive_case},  we proposed the following additive framework to account for the observational noise in the scoring framework:   
\begin{equation*}
\; \left\{ 
	\begin{array}{ll}
		X_{ref} \sim \mathcal{N}(\mu_0,\sigma^2_0), \textrm{ $\mu_{0}$ and $\sigma_{0}$ retrieved from joint model}\\
		Y_{obs} = X_{ref}   +   \epsilon_{obs}, \textrm{ with } \epsilon_{obs} \sim \mathcal{N}(0,0.25), 
	\end{array}
	\right.
\end{equation*}
where  $\mu_{0}=2.55$ and $\sigma_{0}=1.23$ from the fitted distribution in \citep{bessac2018}.    
In Table \ref{tab:score_wind}, log-scores and CRPS are computed as average over the studied time series in the additive Gaussian case with previously given formula in Section \ref{sec:additive_case}. 
The  variance associated with each average score is provided in parenthesis. 
Table \ref{tab:score_wind} shows a significant decrease of the variance when the proposed score is used compared the commonly used in practice score that does not account for measurement error.
One can notice that the variance of the scores used in practice are considerably high limiting the reliability of these computed scores for decision-making purposes. 
Additionally, the new mean scores are closer to the ideal mean log-score and CRPS, showing the benefit of accounting for observational errors in the scoring framework.  
\begin{table}
\centering
\begin{tabular}{|cc|c|}
\hline 
& Mean Score & NWP Prediction \\ 
\hline 
\multirow{ 4}{*}{Log-score} & Ideal score $\Esp(s_{0}(f,X))$ & 1.76 \\ 
& $\Esp(s_{0}(f,Y))$ & 1.97 (1.52) \\ 
& $\Esp(s_{\vee}(f,Y))$ & 1.81 (1.15) \\ 
\hline 
\multirow{ 4}{*}{CRPS} & Ideal score $\Esp(c_{0}(f,X))$ & 0.73 \\ 
& $\Esp(c_{0}(f,Y))$ & 0.82 (0.67) \\ 
& $\Esp(c_{\vee}(f,Y))$ & 0.73 (0.48) \\ 
\hline 
\end{tabular}
\caption{Average scores (log-score and CRPS) computed for the predictive distribution of NWP model, the associated standard deviation is given in parenthesis. The statistics are computed over the entire studied time series. 
The mean ideal score $\Esp(s_{0}(f,X))$, the averaged score computed in practice against the measurements $\Esp(s_{0}(f,Y))$, and the proposed score $\Esp(s_{\vee}(f,X))$ embedding the error in the verification data are computed. }\label{tab:score_wind}
\end{table}
Figure \ref{fig:distrib_score_wind} shows the pdf of the scores considered in Table \ref{tab:score_wind}, the skewness and the large dispersion in the upper tail illustrates with wind speed data cases where the mean is potentially a not informative summary statistics of the whole distribution. 
The non-corrected version of the score has a large variance, raising the concern of reliability on scores when computed on error-prone data.  

\section{Discussion}\label{sec:discussion}

We have quantified, in terms of variance and even distribution, the need to account for the error associated with the verification data when evaluating probabilistic forecasts with scores. 
An additive framework and a multiplicative framework have been studied in detail  to account for the error associated with verification data.  
Both setups involve a probabilistic model for the accounted errors and a probabilistic description of the underling non-observed physical process. 
Although we look only at idealized cases where the involved distributions and their parameters are known, this approach enables us to understand the importance of accounting for the error associated with the verification data. 

\paragraph*{Scores with hidden states}
The proposed scoring framework was  applied to standard additive and a multiplicative models  for which, via Bayesian conjugacy,   the expression  of the new scores could be made explicit.  
However, if the prior on $X$ is not conjugate to the distribution of $Y$, one can derive approximate the conditional distribution  $[X|Y=y[$ through sampling algorithms such as Markov Chain Monte Carlo methods. 
Additionally, one can relax the assumption of known parameters for the distributions of $[X]$ and $[Y|X]$ by considering priors on the parameters (e.g. priors on $\mu_0$ and $\sigma_0$).  
In practice, we rely on the idea that the climatology and-or information on measurement errors can help expressing the distribution of $X$ or its priors. 

In the current literature on scoring with uncertain verification data, most proposed works rely on additive errors as in \citep{ciach1999,saetra2004,mittermaier2015}.   
Multiplicative error models for various error representations have been proposed in modeling studies,  such as for precipitation \citep{mcmillan2011rainfall}, but have not  been considered in scoring correction frameworks. 
Furthermore, additive and multiplicative state-space models can be generalized to non-linear and non-Gaussian state-space model specifications, see Chapter 7 of \citep{cressie2015statistics} for a discussion and examples. 
More generally, one could consider the following state-space model specification    
\begin{equation*}
\; \left\{ 
\begin{array}{l}
X = f(\eta)  \\
Y = g(X, \epsilon), \\
 \end{array}
\right.
\end{equation*} 
where $f$ and $g$ are non-linear functions describing the hidden state and the state-observation mapping, $\eta$ and $\epsilon$ are stochastic terms representing respectively the stochasticity of the hidden state and the observational error.  
This generalized specification of the state-observation model could be integrated in future work in the proposed scoring framework via Bayesian specifications of the new score in order to account for prior information on the verification data $Y$ and its uncertainty and priors on the true state $X$.

\paragraph*{Scores as random variables}
Finally, the study raises the important point of investigating the distribution of scores when the verification data is considered to be a random variable. 
Indeed, investigating the means of scores may not provide sufficient information to compare between score discriminative capabilities.  This has been pointed and investigated in \citep{taillardat2019} in the context of evaluating extremes. 
This topic has also been extensively discussed by \cite{bolin2019} in the context of varying predictability that generates non-homogeneity in the score values that is poorly represented by an average.  
One could choose to take into account   the uncertainty associated with the inference of distribution parameters and compute different statistics of the distribution rather than the mean.   
In this case, a Bayesian setup similar to the current work could elegantly integrate the different hierarchies of knowledge, uncertainties and a priori information to generalize the notion of scores.

\section*{Codes and data}
Codes and data used for this study can be found at \url{https://github.com/jbessac/uncertainty_scoring}. 

\section*{Acknowledgements}
We thank Aur{\'e}lien Ribes for his helpful comments and discussions. 
We also thank the Office of Science and Technology of the Embassy of France in the United States, Washington, DC, for supporting our collaboration through the initiative Make Our Planet Great Again. 
The effort of Julie Bessac is based in part on work supported by the U.S. Department of Energy, Office of Science, Office of Advanced Scientific Computing Research (ASCR) under Contract DE-AC02-06CH11347. 
Part of P. Naveau's work was supported by the European   DAMOCLES-COST-ACTION on compound events, 
and he also  benefited from  French national programs, in particular FRAISE-LEFE/INSU,  MELODY-ANR, ANR-11-IDEX-0004 -17-EURE-0006 and ANR T-REX AAP CE40.


\begin{appendices}
\section{Proof of Equation \eqref{eq:score_variance}} \label{app:total_var}
For any random variable, say $U$, its mean can be written conditionally to the random variable $y$ in the following way: 
 $$ \mathbb{E}\left[ U  \right] = \mathbb{E}\left[ \mathbb{E}\left[ U  | Y=y \right]  \right]. $$ 
 In our case, the variable $U= s_o(f,X)$ and $s_{\vee}(f, y) = \mathbb{E}\left[ U  | Y=y \right] $. 
 This gives  $\mathbb{E}\left[ s_{\vee}(f, Y) \right] =  \mathbb{E}\left[ s_o(f,X) \right]$.  
 To show inequality \eqref{eq:score_variance}, we use the classical variance decomposition 
 $$\mathbb{V}\left[ U  \right]  = \mathbb{V}\left[  \mathbb{E}\left[ U  | Y=y \right]   \right] +  \mathbb{E}\left[  \mathbb{V}\left[ U  | Y=y \right]   \right].$$
With our notations, we have 
\begin{eqnarray*}
\mathbb{V}\left[ s_o(f,X)  \right]  &=& \mathbb{V}\left[  \mathbb{E}\left[ s_o(f,X)  | Y=y \right]   \right] +  \mathbb{E}\left[  \mathbb{V}\left[ s_o(f,X)  | Y=y \right]   \right],\\
 &=& \mathbb{V}\left[  s_{\vee}(f, Y)   \right] +  \mbox{ a non-negative term}.
\end{eqnarray*}
This leads to 
\begin{equation*}
\mathbb{V}\left[ s_o(f,X)  \right]  \geq \mathbb{V}\left[  s_{\vee}(f, Y)   \right].
\end{equation*}
 \hfill $\square$

\section{Proof of Equation \eqref{eq:new_logscore_gauss}} \label{app:cond_logscore}
To express the score proposed in \eqref{eq:new_score}, one needs to derive the conditional distribution $[X|Y=y]$ from Model (A). 
More precisely, the Gaussian conditional distribution of $X$ given $Y=y$ is equal to 
$$[ X | Y=y ] \sim \mathcal{N}\left(\bar{y},\frac{\omega^2 \sigma_0^2}{\sigma_0^2+\omega^2}\right),$$
where $\bar{y}$ is a weighted sum that updates the prior information about $X\sim \mathcal{N}(\mu_0,\sigma^2_0)$ with the observation $Y \sim  \mathcal{N}(\mu_0, \sigma^2_0 + \omega^2)$,  
$$\bar{Y} = \frac{\omega^2}{\sigma_0^2+\omega^2} \mu_0 +  \frac{\sigma_0^2}{\sigma_0^2+\omega^2} Y \sim \mathcal{N}\left(\mu_0,\sigma_0^2 \times \frac{\sigma_0^2}{\sigma_0^2+\omega^2}\right).$$ 
Combining this information with Equations \eqref{eq:new_score} and \eqref{eq:logscore_gauss} leads to 
\begin{eqnarray*}
s_{\vee}(f, y) & =  & \log \sigma + \frac{1}{2 \sigma^2}   \left\{ \mathbb{E}\left[ (X- \mu)^2 | Y=y\right]  \right\}+ \frac{1}{2}\log 2 \pi,\\
 		 & =  & \log \sigma + \frac{1}{2 \sigma^2}   \left\{\mathbb{V}[X|Y=y] +  \left( \mathbb{E}[X|Y=y] -\mu \right)^2 \right\}+ \frac{1}{2}\log 2 \pi,\\
 		 & =  & \log \sigma + \frac{1}{2 \sigma^2}   \left\{\frac{\omega^2\sigma_0^2}{\sigma_0^2+\omega^2}  +  \left( \frac{\omega^2}{\sigma_0^2+\omega^2} \mu_0 +  \frac{\sigma_0^2}{\sigma_0^2+\omega^2} y -\mu \right)^2 \right\}+ \frac{1}{2}\log 2 \pi.\\
\end{eqnarray*} 
By construction, we have 
$$\mathbb{E}_{Y}\left(  s_{\vee}(f, Y) \right) = \mathbb{E}_{\bar{Y}}\left(  s_{\vee}(f, \bar{Y}) \right)  = \mathbb{E}_{X}\left(  s_0(f, X) \right). $$
This means that, to obtain the right score value,  we can first compute $\bar{Y}$ as the best estimator of the unobserved $X$ and then use it into in the corrected score $s_{\vee}(f, \bar{Y})$.  \hfill $\square$

 \section{Proof of Equation \eqref{eq:CRPS_corr}}  \label{app:cond_crps}
To compute the corrected CRPS, one needs to calculate the conditional expectation of $c_0(f,X)$ under the distribution of $[X|Y=y]$. 
We first compute the expectation $E(c_0(f,X))$ and  then substitute $X$ by $\bar{Y}$ and its distribution with mean $a=\bar{y}$ and standard deviation $b=\sqrt{\frac{\omega^2 \sigma_0^2}{\sigma_0^2+\omega^2}}$.
 From Equation \eqref{eq:CRPS_Gauss}  we obtain
\begin{eqnarray}
\nonumber \Esp(c_0(f,X)) = \Esp(X) + 2 \sigma \left[  \Esp\left(\phi\left( \frac{X-\mu}{\sigma}\right)\right) -   \Esp\left(\frac{X-\mu}{\sigma}\overline{\Phi_{0}}\left( \frac{X-\mu}{\sigma}\right)\right)     \right]  - \left[ \mu  + \frac{\sigma}{\sqrt{\pi}}     \right]. 
\end{eqnarray}
If $X$ follows a normal distribution with mean $a$ and variance $b^2$, that is, $X= a + b Z$ with $Z$ a standard random variable, then 
we can define the continuous function 
$h(z) = \overline{\Phi}\left( \frac{a + b z-\mu}{\sigma}\right), $
with 
$h'(z) =-\frac{b}{\sigma} \phi\left( \frac{a + b z-\mu}{\sigma}\right).$
Then, we apply Stein's lemma \citep{Stein81}, which states $\mathbb{E} \left[  h'(Z)  \right] = \mathbb{E} \left[ Z  h(Z)  \right],$
because $Z$ is a standard random variable.
It follows with  the notations $t =\frac{b^2}{2 \sigma^2} $ and $ \lambda = \frac{a-\mu}{\sigma}$ that 
\begin{eqnarray*}
\nonumber \mathbb{E} \left[  \frac{X-\mu}{\sigma}\overline{\Phi}\left( \frac{X-\mu}{\sigma}\right) \right]
\nonumber 	& = & \lambda \Esp \left[  \overline{\Phi}\left( \lambda + \frac{b}{\sigma}Z\right) \right]  + \frac{b}{\sigma} \mathbb{E} \left[ Z  h(Z)  \right],\\ 
\nonumber 	& = & \lambda \Esp \left(\mathbb{P} \left[  Z' > \left( \lambda + \frac{b}{\sigma}Z\right) \right]\right)  + \frac{b}{\sigma}\mathbb{E} \left[  h'(Z)  \right], \\
\nonumber 	& & \textrm{ where $Z'$ has a standard normal distribution}\\ 
\nonumber 	& = & \lambda \Esp \left(\mathbb{P} \left[  Z' -  \frac{b}{\sigma}Z > \lambda \right] \right)  - \frac{b^2}{\sigma^2}  \mathbb{E} \left[  \phi\left( \frac{a + b Z-\mu}{\sigma}\right)  \right],\\ 
\end{eqnarray*} 
with 
\begin{eqnarray*}	
\nonumber \lambda \mathbb{P} \left[ Z' - \frac{b}{\sigma}Z > \lambda \right]	& = & \lambda \overline{\Phi} \left[ \frac{\lambda}{\sqrt{1+b^2/\sigma^2}} \right]  =  \lambda \overline{\Phi} \left[ \frac{a -\mu}{\sqrt{\sigma^2+b^2}} \right]  \\ 
		& = & \frac{a-\mu}{\sigma} \overline{\Phi} \left[ \frac{a -\mu}{\sqrt{\sigma^2+b^2}}   \right]  . \\ 
\end{eqnarray*}	
Then 
\begin{eqnarray*} 
\nonumber \mathbb{E} \left[  \frac{X-\mu}{\sigma}\overline{\Phi}\left( \frac{X-\mu}{\sigma}\right) \right] & = & \frac{a-\mu}{\sigma} \overline{\Phi} \left[ \frac{a -\mu}{\sqrt{\sigma^2+b^2}}   \right] - \frac{b^2}{\sigma^2}  \mathbb{E} \left[  \phi\left( \frac{a + b Z-\mu}{\sigma}\right)  \right],
\end{eqnarray*}
and 
\begin{eqnarray*}
\nonumber \mathbb{E} \left[  \phi\left( \frac{a + b Z-\mu}{\sigma}\right)  \right] & = & \frac{1}{\sqrt{2\pi}} \Esp \left( \exp \left(-\frac{1}{2}\left(\frac{a+bZ-\mu}{\sigma}\right)^2 \right)\right) \\
\nonumber & = & \frac{1}{\sqrt{2\pi}} \Esp \left( \exp \left(-\frac{b^2}{2\sigma^2}\left(Z+\frac{a-\mu}{b}\right)^2 \right)\right) \\
\end{eqnarray*}
$\left(Z+\frac{(a -\mu)}{b}\right)^2$ is a noncentered chi-square distribution with one degree of freedom and a noncentral parameter $(\frac{a-\mu}{b})^2$ with known moment generating function 
$$G\left(t;k=1,\lambda=\left(\frac{a-\mu}{b}\right)^2 \right) = \frac{\exp(\frac{\lambda t}{1-2t})}{(1-2t)^{k/2}}.$$
It follows that 
\begin{eqnarray*}
 \mathbb{E} \left[  \phi\left( \frac{a + b Z-\mu}{\sigma}\right)  \right] & = & \frac{1}{\sqrt{2\pi}} G\left(t=\frac{-b^2}{2 \sigma^2};k=1,\lambda=\left(\frac{a-\mu}{b}\right)^2 \right)  \\
 & = & \frac{1}{\sqrt{2\pi}} \frac{\exp \left(\frac{\frac{-(a-\mu)^2}{b^2} \frac{b^2}{2\sigma^2}}{1+\frac{b^2}{\sigma^2}} \right)}{\sqrt{(1+\frac{b^2}{\sigma^2})}} \\ 
 & = & \frac{\sigma}{\sqrt{2\pi}\sqrt{\sigma^2+b^2}} \exp\left( \frac{-(a-\mu)^2}{2(\sigma^2+b^2)}\right). 
 \end{eqnarray*}
 We obtain 
\begin{eqnarray*}\label{eq:crps_mean}
\nonumber E(c_{0}(f,X)) & = & E(X) + 2 \sigma \left[  \left(1+\frac{b^2}{\sigma^2}\right)E\left(\phi\left( \frac{X-\mu}{\sigma}\right)\right) - \frac{a-\mu}{\sigma} \overline{\Phi} \left[    \frac{a -\mu}{\sqrt{\sigma^2+b^2}}   \right]  \right] - \left(\mu  + \frac{\sigma}{\sqrt{\pi}} \right) \\
\nonumber & = & E(X) + 2 \sigma \left[  \left(1+\frac{b^2}{\sigma^2}\right) \frac{\sigma}{\sqrt{2 \pi (\sigma^2 + b^2)}}  \exp{\left( \frac{-(a-\mu)^{2}}{2(\sigma^2 + b^2)}  \right)}  - \frac{a-\mu}{\sigma} \overline{\Phi} \left[    \frac{a -\mu}{\sqrt{\sigma^2+b^2}}   \right] \right]  \\  
\nonumber && - \left(\mu  + \frac{\sigma}{\sqrt{\pi}} \right) \\
\nonumber & = & E(X) + 2  \left[ \frac{\sqrt{\sigma^2 + b^2}}{\sqrt{2 \pi}}  \exp{\left( \frac{-(a-\mu)^{2}}{2 (\sigma^2 + b^2)}  \right)}  - (a-\mu) \overline{\Phi} \left[    \frac{a -\mu}{\sqrt{\sigma^2+b^2}}   \right] \right]  \\  
&& - \left(\mu  + \frac{\sigma}{\sqrt{\pi}} \right). 
\end{eqnarray*}   
 The expression of \eqref{eq:CRPS_corr} is obtained by substituting $X$ by $\bar{Y}$ and its Gaussian distribution with mean $a=\bar{y}$ and standard deviation $b=\sqrt{\frac{\omega^2 \sigma_0^2}{\sigma_0^2+\omega^2}}$ in the expression \eqref{eq:crps}. This gives 
\begin{eqnarray*}
 c_{\vee}(f,\bar{y}) & = & E(c_{0}(f,X)|Y=y) \\
& = &  \bar{y} - \left( \mu + \frac{\sigma}{\sqrt{\pi}}\right)  \\
&& + 2\left( \frac{\sqrt{\sigma^2 +\frac{\sigma_{0}^{2}\omega^{2}}{\sigma_{0}^{2}+\omega^{2}}}}{\sqrt{2\pi}} \exp\left(-\frac{(\bar{y}-\mu)^2}{2(\sigma^{2}+\frac{\omega^2 \sigma_0^2}{\sigma_0^2+\omega^2})}\right) - (\bar{y} - \mu)\Phi\left( \frac{\bar{y}-\mu}{\sqrt{\sigma^{2}+\frac{\omega^2 \sigma_0^2}{\sigma_0^2+\omega^2}}}\right) \right). 
\end{eqnarray*}
 \hfill $\square$

\section{Proof of Equation \eqref{eq:logSc_distrib}} \label{app:logsc_distrib}
For Model (A), both random variables $Y$ and $\bar{Y}=\frac{\omega^2}{\sigma_0^2+\omega^2} \mu_0 +  \frac{\sigma_0^2}{\sigma_0^2+\omega^2} Y$ are  normally distributed with the same mean $\mu_0$ but different  variances,  
$\sigma_0^2+\omega^2$
and 
$\left( \frac{\sigma_0^2}{\sigma_0^2+\omega^2} \right)^2 ( \sigma_0^2+\omega^2)$, respectively. 
Since a chi-square distribution can be defined as the square of a Gaussian random variable, it follows from Eqn.  \eqref{eq:new_logscore_gauss} that 
$$ s_{0}\left(f, Y\right) \stackrel{d}{=}  a_{0}  +    b_{0}  \chi^2_{0} \mbox{ and } s_{\vee}\left(f, Y\right) \stackrel{d}{=}  a_{\vee}  +    b_{\vee}  \chi^2_{\vee} ,$$
where $\stackrel{d}{=} $ means equality in distribution and 
$$a_{0} = \log \sigma + \frac{1}{2}\log 2 \pi
 \quad \mbox{ and } \quad 
 b_{0} = \frac{\sigma^2_0+\omega^2}{2 \sigma^2} ,$$
 
$$a_{\vee} = \log \sigma + \frac{1}{2 \sigma^2}   \frac{\omega^2\sigma_0^2}{\sigma_0^2+\omega^2} + \frac{1}{2}\log 2 \pi
 \quad \mbox{ and } \quad 
 b_{\vee} = \frac{\sigma^2_0+\omega^2}{2 \sigma^2}  \left( \frac{\sigma_0^2}{\sigma_0^2+\omega^2} \right)^2,$$
with  
$ \chi^2_{0}$ and $ \chi^2_{\vee}$ representing noncentral chi-squared random variable with one degree of freedom and noncentrality parameter
 $$\lambda_{0} = \frac{(\mu_0-\mu)^2}{\sigma_0^2+\omega^2}   \mbox{ and } \lambda_{\vee} = \frac{(\mu_0-\mu)^2}{\sigma_0^2+\omega^2}  \left( \frac{\sigma_0^2+\omega^2}{\sigma_0^2} \right)^2.$$
 \hfill $\square$

 \section{Proof of Equation \eqref{eq:logScCorrGamma}}  \label{app:logsc_mult}
In Model (B), the basic conjugate prior properties of such gamma and inverse gamma distributions  allow us to say that
$[X|Y=y]$ now follows a gamma distribution with parameters $\alpha_0+a$ and $\beta_0+b/y$
\begin{equation}
\nonumber f_{X|Y=y}(x|y)= \frac{(\beta_0+b/y)^{\alpha_0+a}}{\Gamma(\alpha_0+a)} x^{\alpha_0+a-1} \exp\left( - x(\beta_0 + b/y) \right) \mbox{, for $x >0$.}
\end{equation}
It follows that the proposed corrected score is 
\begin{eqnarray*}
\nonumber s_{\vee}(f, y) & = & \mathbb{E}\left[ s_o(f,X) | Y=y\right] \\ 
\nonumber & = & (1-\alpha) \mathbb{E}\left[  \log(X)  | Y=y\right]  + \beta \mathbb{E}\left[ X  | Y=y\right] - \alpha \log\beta + \log \Gamma(\alpha), \\
\nonumber  &=& (1-\alpha) \left( \psi(\alpha_0+a)-\log \left(\beta_0+\frac{b}{y}\right) \right)  + \beta \frac{\alpha_0+a}{\beta_0+\frac{b}{y}} - \alpha \log\beta + \log \Gamma(\alpha). 
\end{eqnarray*}
Indeed $\mathbb{E}\left[ X | Y=y\right]=\frac{\alpha_0+a}{\beta_0+\frac{b}{y}}$ and $\mathbb{E}\left[  \log(X) | Y=y\right] = \psi(\alpha_0+a)-\log \left(\beta_0+\frac{b}{y}\right)$  
where $\psi(x)$ represents  the digamma function defined as the logarithmic derivative of the gamma function, namely, $\psi(x)=d \log \Gamma(x)/dx$.
 \hfill $\square$

 \section{Proof of Equation \eqref{eq:crps_Gamma_corr}} \label{app:crps_mult}
From Equation \eqref{eq:crps_Gamma} we obtain
\begin{eqnarray*}
\nonumber c_{\vee}(f, y) & = & \mathbb{E}\left[ X | Y=y\right] - \left[ \frac{\alpha}{\beta}   +  \frac{1}{\beta B(.5,\alpha)}   \right]  + 2 \left[ \Esp\left[\frac{X}{\beta} f\left( X\right)|Y=y\right] +  \Esp\left[\left(\frac{\alpha}{\beta}-X\right)\overline{F}\left( X\right)|Y=y\right]     \right]. 
\end{eqnarray*}
Since the conditional distribution of $[X|Y=y]$ is known: 
\begin{eqnarray*}
\mathbb{E}\left[\frac{X}{\beta}f(X)|Y=y\right] & = & \frac{1}{\beta}\int_{0}^{+\infty} x f(x) \frac{(\beta_0+b/y)^{\alpha_0+a}}{\Gamma(\alpha_0+a)} x^{\alpha_0+a-1} \exp\left( - x(\beta_0 + b/y) \right) \mathrm{d}x \\
& = &  \frac{\beta^{\alpha-1}(\beta_{0}+b/y)^{\alpha_{0}+a}}{\Gamma(\alpha)\Gamma(\alpha_{0}+a)}\int_{0}^{+\infty} x^{\alpha+\alpha_{0}+a-1} \exp(-(\beta+\beta_{0}+b/y)x) \mathrm{d}x \\
& = & \frac{\beta^{\alpha-1}(\beta_{0}+b/y)^{\alpha_{0}+a}}{\Gamma(\alpha)\Gamma(\alpha_{0}+a)} \frac{1}{(\beta + \beta_{0} + b/y)^{\alpha+\alpha_{0} + a}} \int_{0}^{+\infty} u^{\alpha+\alpha_{0}+a-1} \exp(-u) \mathrm{d}u \\
& = & \frac{\beta^{\alpha-1}(\beta_{0}+b/y)^{\alpha_{0}+a}}{\Gamma(\alpha)\Gamma(\alpha_{0}+a)} \frac{\Gamma(\alpha+\alpha_{0}+a)}{(\beta+ \beta_{0}  + b/y)^{\alpha+\alpha_{0} + a}} \\
& = & \frac{\beta^{\alpha-1}(\beta_{0}+b/y)^{\alpha_{0}+a}}{B(\alpha,\alpha_{0}+a)(\beta+ \beta_{0}  + b/y)^{\alpha+\alpha_{0} + a}},  
\end{eqnarray*}
and the last term 
\begin{eqnarray*}
\mathbb{E}\left(\left(\frac{\alpha}{\beta}-X\right)\bar{F}(X)|Y=y\right)&=&\frac{(\beta_{0}+b/y)^{\alpha_{0}+a}}{\Gamma(\alpha_{0}+a)}\int_{0}^{+\infty} \left( \frac{\alpha}{\beta} - x\right)\left(\int_{x}^{+\infty}\frac{\beta^{\alpha}}{\Gamma(\alpha)}u^{\alpha-1}\exp(-\beta u)\mathrm{d}u\right) \\ 
&& {\small \times} x^{\alpha_{0}+a-1} \exp(-(\beta_{0}+b/y)x) \mathrm{d}x \\
&& \textrm{with } \int_{x}^{+\infty}\frac{\beta^{\alpha}}{\Gamma(\alpha)}u^{\alpha-1}\exp(-\beta u)\mathrm{d}u = \frac{1}{\Gamma(\alpha)} \int_{\beta x}^{+\infty}v^{\alpha-1}\exp(-v)\mathrm{d}v\\
 &=&  \frac{(\beta_{0}+b/y)^{\alpha_{0}+a}}{\Gamma(\alpha)\Gamma(\alpha_{0}+a)}\int_{0}^{+\infty} \left( \frac{\alpha}{\beta} - x\right)\Gamma(\alpha,\beta x) x^{\alpha_{0}+a-1} \exp(-(\beta_{0}+b/y)x) \mathrm{d}x, \\
\end{eqnarray*}
where $\displaystyle \Gamma(\alpha,\beta x)= \int_{\beta x}^{+\infty}v^{\alpha-1}\exp(-v)\mathrm{d}v$ is the upper incomplete gamma function. 
The entire expression of the corrected CRPS expresses as 
\begin{eqnarray}\label{eq:crpsGamma2}
\nonumber c_{\vee}(f, y) & = & \left[ \frac{\alpha}{\beta}   -  \frac{1}{\beta B(.5,\alpha)}   \right] -  \frac{\alpha_{0}+a}{\beta_{0}+\frac{b}{y}} +  2\frac{\beta^{\alpha-1}(\beta_{0}+b/y)^{\alpha_{0}+a}}{B(\alpha,\alpha_{0}+a)(\beta+ \beta_{0}  + b/y)^{\alpha+\alpha_{0} + a}} \\
\nonumber  & + & 2\frac{(\beta_{0}+b/y)^{\alpha_{0}+a}}{\Gamma(\alpha)\Gamma(\alpha_{0}+a)}\int_{0}^{+\infty} \left( \frac{\alpha}{\beta} - x\right)\Gamma(\alpha,\beta x) x^{\alpha_{0}+a-1} \exp(-(\beta_{0}+b/y)x) \mathrm{d}x. 
\end{eqnarray} \hfill $\square$

\section{Additional results: distance between scores distributions}\label{sec:score_wasserstein}
\begin{figure}
    \centering
    \includegraphics[scale=.43]{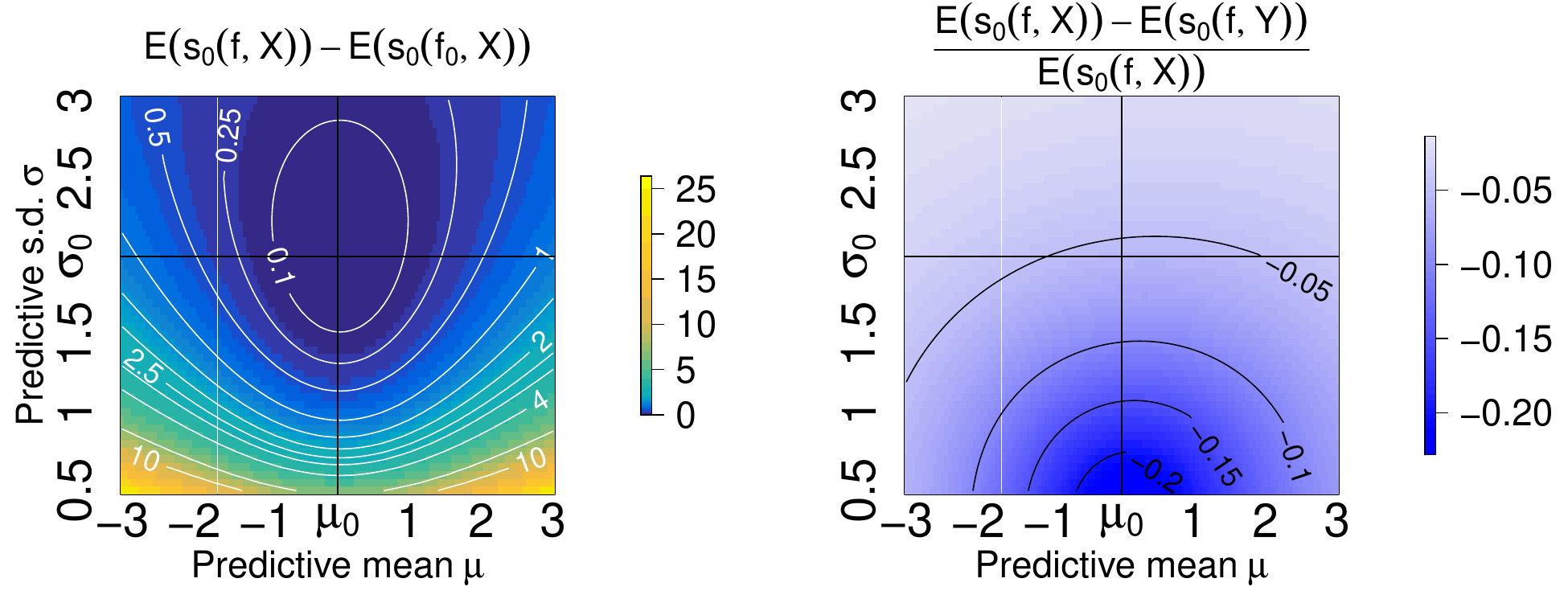}
       \caption{Left: Mean log-score for an imperfect forecast $f$ minus the mean log-score of the perfect forecast $f_0$ when evaluated against perfect data $X$, the imperfect forecast $f \sim \mathcal{N}(\mu,\sigma)$ has varying mean $\mu$ (x-axis) and varying standard deviation $\sigma$ (y-axis). Right: Relative difference between $\Esp(s_{0}(f,X))$ and  $\Esp(s_{0}(f,Y))$. }
    \label{fig:score_surface}
\end{figure}

In order to further study the impact of imperfect verification data and to take full advantages of the score distributions we compute the Wasserstein distance \citep{muskulus2011,santambrogio2015,robin2017} between several scores distributions and compare it to the commonly used score average. 
In particular, through their full distributions we investigate the discriminative skills of scores compared to the use of their mean only.  
The $p$-Wasserstein distance between two probability measures $P$ and $Q$ on $\mathbb{R^d}$ with finite $p$-moments is given by
${\displaystyle W_{p}(P ,Q):=\left(\inf _{\gamma \in \Gamma (P ,Q)}\int _{M\times M}d(x,y)^{p}\,\mathrm {d} \gamma (x,y)\right)^{1/p},}$
where $\Gamma (P ,Q)$ is the set of all joint probability measures on $\mathbb{R^d} \times \mathbb{R^d}$ whose marginals are
$P$ and $Q$. 
In the one-dimensional case as here, the Wasserstein distance can be computed as 
$\displaystyle  W_p(F,G) = \left( \int_0^1 |F^{-1}(u)-G^{-1}(u)|^p du \right)^{1/p},$
with $F$ and $G$ the cdf of $P$ and $Q$ to be compared, $F^{-1}$ and $G^{-1}$ their generalized inverse (or quantile function) and in our case $p=1$. 
The R-package \rm{transport} \citep{Schuhmacher2020} is used to compute Wasserstein distances.   
Figure \ref{fig:score_surface} shows the mean of the log-score minus its minimum $\Esp(s_{0}(f_{0},X))$ and the relative difference between the ideal mean log-score and the mean log-score evaluated against imperfect verification data. 
One can first observe the flatness of the mean log-score around its minimum indicating a lack of discriminative skills of the score mean when comparing several forecasts. 
Secondly, the discrepancy between the score evaluated against perfect and imperfect verification data indicates the effects of error prone verification data as discussed earlier. 

\begin{figure}
    \centering
    \includegraphics[scale=.45]{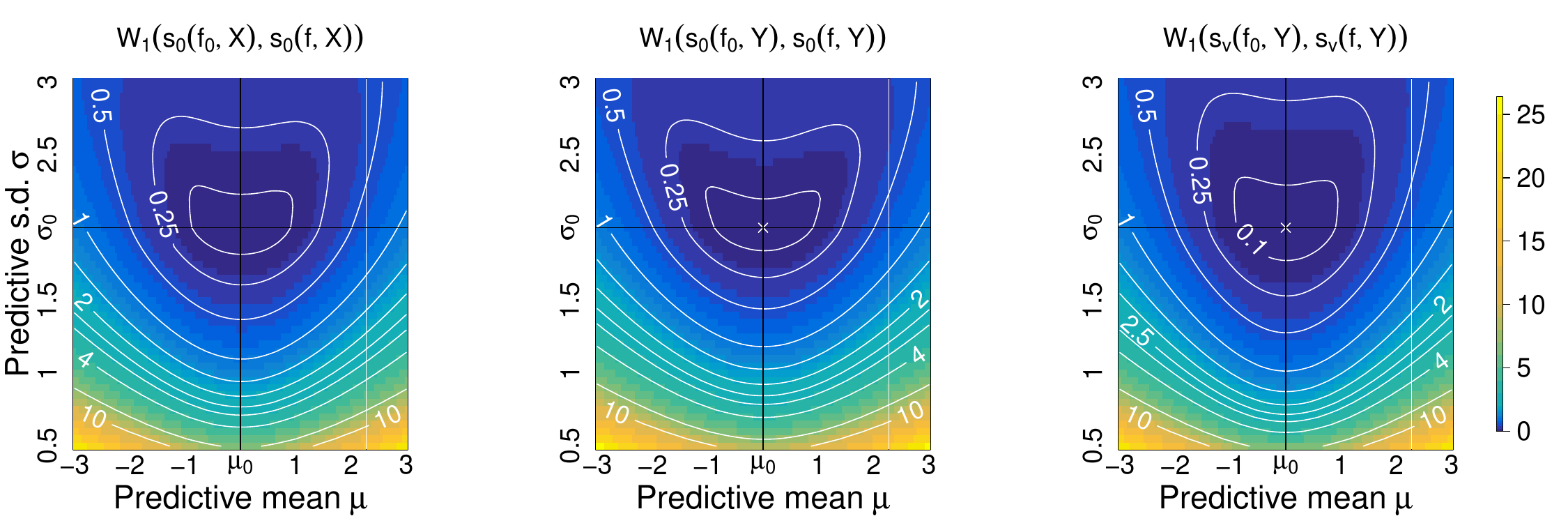}
       \caption{Bottom row: Wasserstein distance $W_{1}(s_{0}(f_{0},.),s_{0}(f,.))$ between log-scores $s_0$ distributions evaluated at the perfect forecast $f_0$ and the imperfect forecast $f$ with varying predictive mean $\mu$ (x-axis) and varying predictive standard deviation $\sigma$ (y-axis).  From left to right, log-scores are evaluated against the hidden true state $X$ via $s_{0}(f,X)$, against $Y$ tainted by observational noise of level $\omega^2=1$ via $s_{0}(f,Y)$ and through the corrected log-score version  $s_{\vee}(f,Y)$.   The verification data $X$ and perfect forecast $f_0$ are distributed according to $f_{0} \sim \mathcal{N}(0,4)$.  
  On the central and right surfaces, the white cross `$\times$' indicates the numerical minimum of each surface. }
    \label{fig:wasserstein}
\end{figure}

Figure \ref{fig:wasserstein} shows the Wasserstein distance between the distributions of three scores evaluated in different contexts (a perfect forecast and an imperfect forecast, true and error-prone verification data).  
Three different log-scores are considered the ideal log-score $s_{0}(.,X)$, the log-score used in practice $s_{0}(.,Y)$ and the proposed corrected score $s_{\vee}(.,Y)$.  
One can notice that Wasserstein distances exhibit stronger gradients (contour lines are narrower) than the mean log-score in Figure \ref{fig:score_surface}.  
In particular, the surfaces delimited by a given contour level are smaller for the proposed score than for the other scores, for instance the area inside the contour of level $z=0.1$ is larger for the mean log-score in Figure \ref{fig:score_surface} than for the Wasserstein distance between score distance.     
This indicates that considering the entire score distribution have the potential to improve the discriminative skills of scoring procedures.  
In particular, imperfect forecasts departing from the perfect forecast will be more sharply discriminated with the Wasserstein distance computed on score distributions.  

Finally, when considering Wasserstein distances associated with the score evaluated on imperfect verification data the minimum of the distances (indicated by white crosses `$\times$' on the central and right panels) is close to the `true' minimum (intersection of $x=\mu_0$ and $y=\sigma_0$).  
This indicates some robustness of the Wasserstein distance between the score distributions when errors are present in the verification data.  
Similar results are obtained for the CRPS and not reported here.  
As stated earlier, developing metrics to express the discriminative skills of a score is beyond the extent of this work.  

\end{appendices}

\bibliographystyle{copernicus}
\bibliography{references.bib}

\begin{thebibliography}{50}
\providecommand{\natexlab}[1]{#1}
\providecommand{\url}[1]{{\tt #1}}
\providecommand{\urlprefix}{URL }
\expandafter\ifx\csname urlstyle\endcsname\relax
  \providecommand{\doi}[1]{https://doi.org/\discretionary{}{}{}#1}\else
  \providecommand{\doi}{https://doi.org/\discretionary{}{}{}\begingroup
  \urlstyle{rm}\Url}\fi

\bibitem[{Anderson(1996)}]{Anderson96}
Anderson, J.~L.: A method for producing and evaluating probabilistic forecasts
  from ensemble model integrations, Journal of Climate, 9, 1518--1530, 1996.

\bibitem[{Bessac et~al.(2018)Bessac, Constantinescu, and Anitescu}]{bessac2018}
Bessac, J., Constantinescu, E., and Anitescu, M.: Stochastic simulation of
  predictive space--time scenarios of wind speed using observations and
  physical model outputs, The Annals of Applied Statistics, 12, 432--458, 2018.

\bibitem[{Bolin and Wallin(2019)}]{bolin2019}
Bolin, D. and Wallin, J.: Scale invariant proper scoring rules Scale
  dependence: Why the average CRPS often is inappropriate for ranking
  probabilistic forecasts, arXiv preprint arXiv:1912.05642, 2019.

\bibitem[{Bowler(2008)}]{Bowler08}
Bowler, N.~E.: Accounting for the effect of observation errors on verification
  of MOGREPS, Meteorological Applications, 15, 199--205, 2008.

\bibitem[{Br{\"o}cker and Ben~Bouall{\`e}gue(2020)}]{Broecker20}
Br{\"o}cker, J. and Ben~Bouall{\`e}gue, Z.: Stratified rank histograms for
  ensemble forecast verification under serial dependence, Quarterly Journal of
  the Royal Meteorological Society, 146, 1976--1990,
  \doi{https://doi.org/10.1002/qj.3778},
  \urlprefix\url{https://rmets.onlinelibrary.wiley.com/doi/abs/10.1002/qj.3778},
  2020.

\bibitem[{Br{\"o}cker and Smith(2007)}]{brocker2007}
Br{\"o}cker, J. and Smith, L.~A.: Scoring probabilistic forecasts: The
  importance of being proper, Weather and Forecasting, 22, 382--388, 2007.

\bibitem[{Candille and Talagrand(2008)}]{Candille08}
Candille, G. and Talagrand, O.: Retracted and replaced: Impact of observational
  error on the validation of ensemble prediction systems, Quarterly Journal of
  the Royal Meteorological Society, 134, 509--521, 2008.

\bibitem[{Ciach and Krajewski(1999)}]{ciach1999}
Ciach, G.~J. and Krajewski, W.~F.: On the estimation of radar rainfall error
  variance, Advances in Water Resources, 22, 585--595, 1999.

\bibitem[{Cressie and Wikle(2015)}]{cressie2015statistics}
Cressie, N. and Wikle, C.~K.: Statistics for spatio-temporal data, John Wiley
  \& Sons, 2015.

\bibitem[{Daley(1993)}]{Daley93}
Daley, R.: Estimating observation error statistics for atmospheric data
  assimilation., Annales Geophysicae, 11, 634--647, 1993.

\bibitem[{Diebold and Mariano(2002)}]{diebold2002}
Diebold, F.~X. and Mariano, R.~S.: Comparing predictive accuracy, Journal of
  Business \& economic statistics, 20, 134--144, 2002.

\bibitem[{Dirkson et~al.(2019)Dirkson, Merryfield, and Monahan}]{dirkson2019}
Dirkson, A., Merryfield, W.~J., and Monahan, A.~H.: Calibrated probabilistic
  forecasts of Arctic sea ice concentration, Journal of Climate, 32,
  1251--1271, 2019.

\bibitem[{Ferro(2017)}]{Ferro17}
Ferro, C. A.~T.: Measuring forecast performance in the presence of observation
  error, Quarterly Journal of the Royal Meteorological Society, 143,
  2665--2676, \doi{10.1002/qj.3115},
  \urlprefix\url{http://dx.doi.org/10.1002/qj.3115}, 2017.

\bibitem[{Gelman et~al.(2013)Gelman, Carlin, Stern, Dunson, Vehtari, and
  Rubin}]{gelman2013}
Gelman, A., Carlin, J.~B., Stern, H.~S., Dunson, D.~B., Vehtari, A., and Rubin,
  D.~B.: Bayesian data analysis, CRC press, 2013.

\bibitem[{Gneiting and Raftery(2007)}]{Gneiting07}
Gneiting, T. and Raftery, A.~E.: Strictly proper scoring rules, prediction, and
  estimation, Journal of the American Statistical Association, 102, 359--378,
  2007.

\bibitem[{Gneiting et~al.(2005)Gneiting, Raftery, Westveld~III, and
  Goldman}]{Gneiting05}
Gneiting, T., Raftery, A.~E., Westveld~III, A.~H., and Goldman, T.: Calibrated
  probabilistic forecasting using ensemble model output statistics and minimum
  {CRPS} estimation, Monthly Weather Review, 133, 1098--1118, 2005.

\bibitem[{Gneiting et~al.(2007)Gneiting, Balabdaoui, and Raftery}]{Gneiting07b}
Gneiting, T., Balabdaoui, F., and Raftery, A.~E.: Probabilistic forecasts,
  calibration and sharpness, Journal of the Royal Statistical Society: Series B
  (Statistical Methodology), 69, 243--268, 2007.

\bibitem[{Gorgas and Dorninger(2012)}]{Gorgas12}
Gorgas, T. and Dorninger, M.: Quantifying verification uncertainty by reference
  data variation, Meteorologische Zeitschrift, 21, 259--277, 2012.

\bibitem[{Hamill(2001)}]{Hamill01}
Hamill, T.~M.: Interpretation of rank histograms for verifying ensemble
  forecasts, Monthly Weather Review, 129, 550--560, 2001.

\bibitem[{Hamill and Juras(2006)}]{hamill2006}
Hamill, T.~M. and Juras, J.: Measuring forecast skill: Is it real skill or is
  it the varying climatology?, Quarterly Journal of the Royal Meteorological
  Society, 132, 2905--2923, 2006.

\bibitem[{Janji{\'c} et~al.(2017)Janji{\'c}, Bormann, Bocquet, Carton, Cohn,
  Dance, Losa, Nichols, Potthast, Waller, and Weston}]{Janjic17}
Janji{\'c}, T., Bormann, N., Bocquet, M., Carton, J.~A., Cohn, S.~E., Dance,
  S.~L., Losa, S.~N., Nichols, N.~K., Potthast, R., Waller, J.~A., and Weston,
  P.: On the representation error in data assimilation, Quarterly Journal of
  the Royal Meteorological Society, 2017.

\bibitem[{Jolliffe(2007)}]{jolliffe2007}
Jolliffe, I.~T.: Uncertainty and inference for verification measures, Weather
  and Forecasting, 22, 637--650, 2007.

\bibitem[{Jolliffe and Stephenson(2004)}]{Jolliffe04}
Jolliffe, T. and Stephenson, D.~B.: Forecast verification: A practitioner's
  guide in atmospheric science. Edited by Ian Wiley, Chichester, 2003. xiv+240
  pp. ISBN 0 471 49759 2, Weather, 59, 132--132, \doi{10.1256/wea.123.03},
  \urlprefix\url{http://dx.doi.org/10.1256/wea.123.03}, 2004.

\bibitem[{Kalman(1960)}]{kalman1960}
Kalman, R.~E.: A new approach to linear prediction and filtering problems,
  Transactions of the ASME--Journal of Basic Engineering, pp. 35--45, 1960.

\bibitem[{Kalman and Bucy(1961)}]{kalman1961}
Kalman, R.~E. and Bucy, R.~S.: New results in linear filtering and prediction
  theory, 1961.

\bibitem[{Kavetski et~al.(2006{\natexlab{a}})Kavetski, Kuczera, and
  Franks}]{kavetski1}
Kavetski, D., Kuczera, G., and Franks, S.~W.: Bayesian analysis of input
  uncertainty in hydrological modeling: 1. Theory, Water Resources Research,
  42, \doi{https://doi.org/10.1029/2005WR004368},
  \urlprefix\url{https://agupubs.onlinelibrary.wiley.com/doi/abs/10.1029/2005WR004368},
  2006{\natexlab{a}}.

\bibitem[{Kavetski et~al.(2006{\natexlab{b}})Kavetski, Kuczera, and
  Franks}]{kavetski2}
Kavetski, D., Kuczera, G., and Franks, S.~W.: Bayesian analysis of input
  uncertainty in hydrological modeling: 2. Application, Water Resources
  Research, 42, \doi{https://doi.org/10.1029/2005WR004376},
  \urlprefix\url{https://agupubs.onlinelibrary.wiley.com/doi/abs/10.1029/2005WR004376},
  2006{\natexlab{b}}.

\bibitem[{Kleen(2019)}]{kleen2019}
Kleen, O.: Measurement Error Sensitivity of Loss Functions for Distribution
  Forecasts, Available at SSRN 3476461, 2019.

\bibitem[{McMillan et~al.(2011)McMillan, Jackson, Clark, Kavetski, and
  Woods}]{mcmillan2011rainfall}
McMillan, H., Jackson, B., Clark, M., Kavetski, D., and Woods, R.: Rainfall
  uncertainty in hydrological modelling: An evaluation of multiplicative error
  models, Journal of Hydrology, 400, 83--94, 2011.

\bibitem[{Mittermaier and Stephenson(2015)}]{mittermaier2015}
Mittermaier, M.~P. and Stephenson, D.~B.: Inherent bounds on forecast accuracy
  due to observation uncertainty caused by temporal sampling, Monthly Weather
  Review, 143, 4236--4243, 2015.

\bibitem[{Murphy(1973)}]{murphy1973}
Murphy, A.~H.: A new vector partition of the probability score, Journal of
  Applied Meteorology, 12, 595--600, 1973.

\bibitem[{Murphy and Winkler(1987)}]{Murphy87}
Murphy, A.~H. and Winkler, R.~L.: A general framework for forecast
  verification, Monthly Weather Review, 115, 1330--1338, 1987.

\bibitem[{Muskulus and Verduyn-Lunel(2011)}]{muskulus2011}
Muskulus, M. and Verduyn-Lunel, S.: Wasserstein distances in the analysis of
  time series and dynamical systems, Physica D: Nonlinear Phenomena, 240,
  45--58, 2011.

\bibitem[{Pappenberger et~al.(2009)Pappenberger, Ghelli, Buizza, and
  Bodis}]{Pappenberger09}
Pappenberger, F., Ghelli, A., Buizza, R., and Bodis, K.: The skill of
  probabilistic precipitation forecasts under observational uncertainties
  within the generalized likelihood uncertainty estimation framework for
  hydrological applications, Journal of Hydrometeorology, 10, 807--819, 2009.

\bibitem[{Pinson and Hagedorn(2012)}]{Pinson12}
Pinson, P. and Hagedorn, R.: Verification of the ECMWF ensemble forecasts of
  wind speed against analyses and observations, Meteorological Applications,
  19, 484--500, 2012.

\bibitem[{Robert and Casella(2013)}]{robert2013}
Robert, C. and Casella, G.: Monte Carlo statistical methods, Springer Science
  \& Business Media, 2013.

\bibitem[{Robin et~al.(2017)Robin, Yiou, and Naveau}]{robin2017}
Robin, Y., Yiou, P., and Naveau, P.: Detecting changes in forced climate
  attractors with {W}asserstein distance, Nonlinear Processes in Geophysics,
  24, 393--405, 2017.

\bibitem[{Saetra et~al.(2004)Saetra, Hersbach, Bidlot, and
  Richardson}]{saetra2004}
Saetra, O., Hersbach, H., Bidlot, J.-R., and Richardson, D.~S.: Effects of
  observation errors on the statistics for ensemble spread and reliability,
  Monthly weather review, 132, 1487--1501, 2004.

\bibitem[{Santambrogio(2015)}]{santambrogio2015}
Santambrogio, F.: Optimal transport for applied mathematicians, Birk{\"a}user,
  NY, 55, 94, 2015.

\bibitem[{Scheuerer and M{\"o}ller(2015)}]{Scheuerer15}
Scheuerer, M. and M{\"o}ller, D.: Probabilistic wind speed forecasting on a
  grid based on ensemble model output statistics, The Annals of Applied
  Statistics, 9, 1328--1349, 2015.

\bibitem[{Schuhmacher et~al.(2020)Schuhmacher, Bähre, Gottschlich, Hartmann,
  Heinemann, Schmitzer, Schrieber, and Wilm}]{Schuhmacher2020}
Schuhmacher, D., Bähre, B., Gottschlich, C., Hartmann, V., Heinemann, F.,
  Schmitzer, B., Schrieber, J., and Wilm, T.: {transport}: Computation of
  Optimal Transport Plans and Wasserstein Distances,
  \urlprefix\url{https://cran.r-project.org/package=transport}, r package
  version 0.12-2, 2020.

\bibitem[{Skamarock et~al.(2008)Skamarock, Klemp, Dudhia, Gill, Barker, Duda,
  Huang, Wang, and Powers}]{Skamarock08}
Skamarock, W., Klemp, J., Dudhia, J., Gill, D., Barker, D., Duda, M., Huang,
  X.-Y., Wang, W., and Powers, J.: A description of the {A}dvanced {R}esearch
  {WRF} Version 3, Tech. Rep. Tech Notes-475+ STR, NCAR, 2008.

\bibitem[{Stein(1981)}]{Stein81}
Stein, C.~M.: Estimation of the mean of a multivariate normal distribution, The
  Annals of Statistics, pp. 1135--1151, 1981.

\bibitem[{Taillardat et~al.(2016)Taillardat, Mestre, Zamo, and
  Naveau}]{Taillardat16}
Taillardat, M., Mestre, O., Zamo, M., and Naveau, P.: Calibrated Ensemble
  Forecasts using Quantile Regression Forests and Ensemble Model Output
  Statistics., Monthly Weather Review, 144, 2375--2393,
  \doi{10.1175/MWR-D-15-0260.1},
  \urlprefix\url{https://doi.org/10.1175/MWR-D-15-0260.1}, 2016.

\bibitem[{Taillardat et~al.(2019)Taillardat, Foug{\`e}res, Naveau, and
  de~Fondeville}]{taillardat2019}
Taillardat, M., Foug{\`e}res, A.-L., Naveau, P., and de~Fondeville, R.: Extreme
  events evaluation using CRPS distributions, arXiv preprint arXiv:1905.04022,
  2019.

\bibitem[{Waller et~al.(2014)Waller, Dance, Lawless, and Nichols}]{Waller14}
Waller, J.~A., Dance, S.~L., Lawless, A.~S., and Nichols, N.~K.: Estimating
  correlated observation error statistics using an ensemble transform Kalman
  filter, Tellus A: Dynamic Meteorology and Oceanography, 66, 23\,294, 2014.

\bibitem[{Weijs and Van De~Giesen(2011)}]{Weijs11}
Weijs, S.~V. and Van De~Giesen, N.: Accounting for observational uncertainty in
  forecast verification: an information-theoretical view on forecasts,
  observations, and truth, Monthly Weather Review, 139, 2156--2162, 2011.

\bibitem[{Weijs et~al.(2010)Weijs, Van~Nooijen, and Van De~Giesen}]{weijs2010}
Weijs, S.~V., Van~Nooijen, R., and Van De~Giesen, N.: Kullback--{L}eibler
  divergence as a forecast skill score with classic
  reliability--resolution--uncertainty decomposition, Monthly Weather Review,
  138, 3387--3399, 2010.

\bibitem[{Wilks(2010)}]{wilks2010}
Wilks, D.~S.: Sampling distributions of the Brier score and Brier skill score
  under serial dependence, Quarterly Journal of the Royal Meteorological
  Society, 136, 2109--2118, 2010.

\bibitem[{Zamo and Naveau(2018)}]{Zamo18}
Zamo, M. and Naveau, P.: Estimation of the Continuous Ranked Probability Score
  with Limited Information and Applications to Ensemble Weather Forecasts,
  Mathematical Geosciences, 50, 209--234, 2018.

\end{thebibliography}

\end{document}